\documentclass[fleqn,usenatbib]{mnras}

\usepackage{newtxtext,newtxmath}
\usepackage[T1]{fontenc}

\DeclareRobustCommand{\VAN}[3]{#2}
\let\VANthebibliography\thebibliography
\def\thebibliography{\DeclareRobustCommand{\VAN}[3]{##3}\VANthebibliography}

\usepackage{main}

\title[The Cooling of Old White Dwarfs in 47 Tucanae]{The Cooling of Old White Dwarfs in 47 Tucanae}

\author[L. Fleury et al.]{
Leesa Fleury\textsuperscript{\Large \orcidlink{0000-0002-7022-8380}},$^{1}$\thanks{E-mail: lfleury@phas.ubc.ca (LF); heyl@phas.ubc.ca (JH)}
Harvey Richer\textsuperscript{\Large \orcidlink{0000-0001-9002-8178}},$^{1}$\thanks{Deceased}
\addtocounter{footnote}{-2}\addtocounter{Hfootnote}{-2}
and Jeremy Heyl\textsuperscript{\Large \orcidlink{0000-0001-9739-367X}}$^{1}$\footnotemark
\\
$^{1}$Department of Physics and Astronomy, University of British Columbia, Vancouver, BC V6T 1Z1, Canada
}

\date{Accepted XXX. Received YYY; in original form ZZZ}

\pubyear{\the\year{}}

\begin{document}
\label{firstpage}
\pagerange{\pageref{firstpage}--\pageref{lastpage}}
\maketitle

% Abstract of the paper
\begin{abstract}
We analyse the cooling of white dwarfs in the globular cluster 47 Tucanae (47 Tuc) using deep observations from the {\it Hubble Space Telescope} that resolve the white dwarf cooling sequence to late enough cooling times that the envelope has become convectively coupled to the core. At these late times, the thickness of the outer H envelope is an important consideration in modelling the cooling. Using the stellar evolution software Modules for Experiments in Stellar Astrophysics, we create a suite of white dwarf cooling models for different thicknesses of the H envelope and different white dwarf masses. An unbinned likelihood analysis is performed to compare the cooling models to the observations in order to constrain the values of these key parameters. We find that thicker H envelopes are preferred, with the best-fitting models reproducing the observed cumulative 47 Tuc white dwarf luminosity functions well.
\end{abstract}

% Select between one and six entries from the list of approved keywords.
% Don't make up new ones.

%%%%%%%%%%%%%%%%%%%%%%%%%%%%%%%%%%%%%%%%%%%%%%%%%%

%%%%%%%%%%%%%%%%% BODY OF PAPER %%%%%%%%%%%%%%%%%%

\section{Introduction} \label{sec:deepacs_intro}

Globular clusters are useful environments for studying white dwarf cooling because they provide populations of stars with many well controlled parameters, such as distance, interstellar reddening, age, and metallicity \citep[e.g.][]{1985ApJS...58..561V,1996AJ....112.1487H,2010RSPTA.368..755K}.
Though metallicity is not directly a concern for observations of white dwarfs, since element sedimentation causes metals to sink below the surface envelope of the white dwarf by the early stages of white dwarf cooling, it affects the age and composition of the white dwarfs through earlier stages of stellar evolution.
For a coeval population of stars with the same initial metallicity, white dwarfs of a similar mass are formed through single stellar evolution at an approximately constant rate.

Unlike open clusters, for which the parameters mentioned above are also well controlled, globular clusters are very old and typically much more rich (i.e. well-populated).
These additional properties of globular clusters enable white dwarf cooling to be studied to much later times and with the greater statistical power of a larger sample size compared to what could be achieved with open clusters.
The globular cluster 47 Tucanae (47 Tuc), in particular, has an especially rich white dwarf population and is known to be very old, with an age of $\sim 10~\mathrm{Gyr}$ \citep{2013Natur.500...51H}.
Furthermore, the distance to 47 Tuc has been well determined \citep{Chen2018}.

47 Tuc is one of the most widely studied globular clusters, and white dwarfs in 47 Tuc have been used to study a variety of aspects of white dwarf cooling.
For example, \citet{Obertas2018} studied the onset of convective coupling and core crystallisation at late times in the white dwarf cooling process by comparing white dwarf evolution models to deep \textit{Hubble Space Telescope} (HST) observations of 47 Tuc white dwarfs.
As another example, HST observations of younger white dwarfs in 47 Tuc were used by \citet{Goldsbury2016} to study white dwarf cooling by the emission of neutrinos and to constrain neutrino physics.
In both of these cases, the typical envelope thickness of white dwarfs in 47 Tuc was an important ancillary parameter in studying the relevant aspects of white dwarf cooling due to the relation between the envelope thickness and the cooling rate.

The envelope thickness is particularly important for the cooling rate at late cooling times when the envelope becomes convectively coupled to the core of the white dwarf.
Convective coupling of the envelope to the core occurs when the convective layer that eventually develops in the outer layers of a white dwarf \citep{1968Ap&SS...2..375B} breaks through to the degenerate interior, which enables energy to be transported from the core to the surface more quickly than if the envelope had remained radiative \citep[for a review, see e.g.][]{Fontaine2001}.
This results in the white dwarf initially appearing more luminous at the onset of convective coupling than it otherwise would have, which manifests as a bump in the cooling curve showing the evolution of luminosity as a function of cooling time.
\citet{1990ApJS...72..335T} argued that the size of this bump depends sensitively on the thickness of the H envelope.

The work of \citet{Obertas2018} used observations that were deep enough to resolve this feature in the cooling curve that becomes important at late cooling times.
The focus of that work was on the theoretical aspects of modelling the freezing of the white dwarfs and the coincident onset of convective coupling.
\citet{Obertas2018} considered models with both convection and freezing, with convection but without freezing, and with neither convection nor freezing.
The models were compared to the data visually through plots and with a simple Kolmogorov-Smirnov (KS) test to verify statistical significance, but the authors did not fit their models to the data or perform a more extensive statistical analysis.
Furthermore, though \citet{Obertas2018} note the thickness of the H envelope as a parameter that can affect the size of the convective bump in the cooling curve, the effect of varying the envelope thickness was not studied in that work; all of the models had the same white dwarf mass and envelope thickness.
\citet{Goldsbury2016} performed a more rigorous statistical analysis with the thickness of the H envelope as one of the model parameters that was varied.
However, the models were compared to data restricted to brighter white dwarfs for which neutrino cooling was an important effect but that were much too young for convective coupling of the envelope to the core to be a relevant effect.

In this work, we analyse the cooling of white dwarfs in 47 Tuc to late cooling times where convective coupling becomes important for the cooling rate.
We perform stellar evolution simulations using the software Modules for Experiments in Stellar Astrophysics \citep[MESA;][]{mesa1,mesa2,mesa3,mesa4,mesa5,mesa6} to create white dwarf cooling models for different envelope thicknesses and white dwarf masses.
We compare these cooling models to the deep HST data considered by \citet{Obertas2018} using an unbinned likelihood analysis procedure similar to that of \citet{Goldsbury2016}.

% \clearpage
\section{Data} \label{sec:deepacs_data}

Our data consists of archival HST deep observations of the outer field of 47 Tuc imaged by the Advanced Camera for Surveys (ACS) using the Wide Field Channel (WFC), as described by \citet{2012AJ....143...11K}. 
These data were collected over 121 orbits during the time period extending from January 2010 to October 2010 as part of the HST Cycle 17 proposal GO-11677 (PI: H. Richer).
The ACS/WFC observations were done using the broadband filters F606W and F814W. 
The deep exposures had a total integrated exposure time of 163.7~ks across 117 exposures in F606W and 172.8~ks across 125 exposures in F814W.
These observations were centred at sky coordinates of $\alpha = \racoord{00}{22}{39}$ and $\delta = \deccoord{-72}{04}{04}$ in the international celestial reference frame at the reference epoch J2000, where $\alpha$ is the right ascension and $\delta$ is the declination. 
This corresponds to a distance of about 6.7~arcminutes (8.8~pc) from the cluster centre, which is located at $\alpha = \racoord{00}{24}{05.71}$ and $\delta = \deccoord{-72}{04}{52.7}$ \citep{2010AJ....140.1830G}.

For each filter, images from the various exposures were combined into a single final, stacked image. Photometric, astrometric, and morphological measurements were then performed on the final, stacked images for the two filters using iterative point-spread function (PSF) fitting techniques. 
Morphological information from the PSF fitting is stored in the \sharpparam\ diagnostic parameter, which provides a measure of how much broader the source's profile is compared to the PSF profile it was fitted to. 
The \sharpparam\ parameter provides a way to distinguish stars from other contaminant sources such as galaxies and cosmic rays.

The PSF fitting produced a catalogue of sources that contains the F606W and F814W magnitudes, position, chi goodness-of-fit statistic, and \sharpparam\ statistic determined for each source, with the magnitudes reported in the Vega magnitude system.
The full image processing and PSF fitting procedures are described in further detail in \citet{2012AJ....143...11K}, and the resultant catalogue is publicly available through the Mikulski Archive for Space Telescopes (MAST) as a High-Level Science Product (HLSP)%
\footnote{\label{footnote:deepACS}%
The final stacked images, source catalogue, and artificial stars data are available at \url{https://archive.stsci.edu/prepds/deep47tuc/}.}.
In addition to this catalogue, the results of the artificial stars tests documented in \citet{2012AJ....143...11K,2013ApJ...763..110K} are also publicly available as part of the same HLSP collection.
These artificial stars tests are used to characterise the photometric uncertainties and completeness of the data in the final catalogue, and are discussed in detail in \cref{sec:deepacs_artificial_stars}.

Photometric observations of 47 Tuc white dwarfs are contaminated at the faint end of the white dwarf cooling sequence by the Small Magellanic Cloud (SMC).
Though the main body of the SMC lies more than $2^\circ$ away from 47 Tuc%
\footnote{\label{footnote:SMC_coords}The SMC is located at sky coordinates $\alpha = \racoord{00}{52}{45}$, $\delta = \deccoord{-72}{49}{43}$ \citep{2019ApJS..245...25J}. This is an angular distance of $\sim 2.28^\circ$ from the centre of 47 Tuc and $\sim 2.39^\circ$ from the centre of the ACS/WFC observing field.}, %
a diffuse population of SMC stars persists out to very large radii and is present in the background of our 47 Tuc observations.
This background SMC population overlaps with the faint end of the 47 Tuc white dwarf cooling sequence in the colour-magnitude diagram (CMD) of our data, which is the region of the cooling sequence we are most interested in.
Fortunately, the SMC is moving with respect to 47 Tuc fast enough that the two populations can be mostly separated in proper motion space, and thus most of the SMC contaminants can be removed from our data.
The reduction procedures to determine the proper motions accompanying the ACS/WFC photometric observations are described in detail in \citet{2013ApJ...771L..15R}.
Note that we are interested in objects much fainter than those considered in \citet{2013ApJ...771L..15R} and that the uncertainty in the proper motions increases with magnitude.
Our proper motion data thus has larger uncertainties and appears more dispersed overall than that of \citet{2013ApJ...771L..15R}.
Since we are interested in the white dwarf cooling sequence down to very faint magnitudes, it is important to include the proper motions for the fainter white dwarfs, even though it results in a sample with larger proper motion uncertainties.

We identify white dwarfs associated with 47 Tuc in the HST data by making cuts in \sharpparam, proper motion, and the CMD. 
The cut in \sharpparam\ enables us to clean the data by removing contaminants such as cosmic rays and galaxies whose photometric profiles do not match the expected PSF profile for stars.
The cut in proper motion is used to select sources likely to be members of 47 Tuc, further cleaning the data by removing most of the SMC and field stars. 
For this data cleaning, we perform a \sharpparam\ cut that selects objects with $\lvert \text{\sharpparam} \rvert < 0.5$ and a proper motion cut that selects objects with a total proper motion $< 2.5~\textrm{mas}~\textrm{yr}^{-1}$ relative to the mean proper motion of 47 Tuc. 
After cleaning the data using these cuts in \sharpparam\ and proper motion, we then perform a cut in the CMD to select our white dwarf sample, and the boundaries of this cut in the CMD define the data space for the unbinned likelihood analysis. 
The data cleaning procedure, including the choice of which cuts to make, is further explained in detail in \cref{sec:deepacs_data_cleaning}. 
The unbinned likelihood analysis, including the CMD data space selection, is discussed in \cref{sec:deepacs_unbinned_likelihood}.

\section{Artificial Stars Tests} \label{sec:deepacs_artificial_stars}

The artificial stars tests were performed by adding artificial sources to the final stacked images and running the new images through the same PSF fitting procedure as was used for the real data.
Each artificial source was given a unique ID in order to track whether the artificial source was detected by the PSF fitting procedure and, if the source was detected, to compare the output values of its magnitude and position to the known true value.
The artificial stars tests and the corresponding artificial sources catalogue are described in detail in \citet{2012AJ....143...11K}.

The catalogue of artificial sources produced by these artificial stars tests was used to construct a photometric error distribution function in a procedure similar to what is described in \citet{Goldsbury2016}.
Let $\mathrm{F606W}_\mathrm{in}$ and $\mathrm{F814W}_\mathrm{in}$ denote the input magnitudes of a source, and let $\mathrm{F606W}_\mathrm{out}$ and $\mathrm{F814W}_\mathrm{out}$ denote the output magnitudes determined by the PSF fitting procedure.
The photometric error distribution function
\begin{equation}
    E = E\left(\Delta \mathrm{F606W}, \Delta \mathrm{F814W}; \mathrm{F606W}_\mathrm{in}, \mathrm{F814W}_\mathrm{in}\right)
    \label{eq:deepacs_errdist}
\end{equation}
gives the joint probability density, normalised to the number of input stars, of the magnitude errors $\Delta \mathrm{F606W}$ and $\Delta \mathrm{F814W}$ as a function parametrized by the input magnitudes, where the errors are quantified as the differences between the output and input values of the magnitudes, $\Delta \mathrm{F606W} = \mathrm{F606W}_\mathrm{out} - \mathrm{F606W}_\mathrm{in}$ and $\Delta \mathrm{F814W} = \mathrm{F814W}_\mathrm{out} - \mathrm{F814W}_\mathrm{in}$.
The normalisation of the error distribution function as a function of the input magnitudes is simply the completeness,
\begin{equation}
    C\left(\mathrm{F606W}_\mathrm{in}, \mathrm{F814W}_\mathrm{in}\right) = \iint\limits_{-\infty}^{+\infty} E \, \mathrm{d}\left(\Delta \mathrm{F606W}\right) \, \mathrm{d}\left(\Delta \mathrm{F814W}\right),
\end{equation}
which in general is less than unity because not all of the sources that are actually present are recovered by the photometric reduction procedure.
The completeness quantifies the probability of detecting a source and can in principle take values in the range of 0 to 1, though in practice the observations become unusable if the completeness becomes too poor.

In general, the error distribution function and completeness also depend on the position from the centre of the cluster.
However, for the data considered in this work, the position dependence is negligible.
Furthermore, it should be noted that since all of the cooling models that will be considered in our analysis lie along approximately the same curve in colour-magnitude (and likewise magnitude-magnitude) space before accounting for photometric errors, the photometric error distribution only needs to be constructed at the combination of $\left(\mathrm{F606W}_\mathrm{in}, \mathrm{F814W}_\mathrm{in}\right)$ values that lie along this sequence.

\section{Data Cleaning Procedures} \label{sec:deepacs_data_cleaning}

\subsection{Overview} \label{sec:deepacs_overview_data_cleaning}

To clean our deep HST ACS/WFC data, we want to remove sources in the catalogue that (1) are not stars or (2) are not members of 47 Tuc.
In the latter case, we are particularly concerned with removing SMC stars that contaminate the faint end of the 47 Tuc white dwarf cooling sequence in the CMD.
A cut in \sharpparam\ allows us to remove objects that are not stars, while a cut in proper motion allows us to remove objects that are unlikely to be 47 Tuc cluster members.
We calibrate our data cleaning procedure using 47 Tuc main-sequence stars to choose what cuts to make in \sharpparam\ and proper motion and to quantify the effect of these cuts on the completeness of our white dwarf sample.
We furthermore quantify the number of SMC stars expected to survive the cleaning procedure and contaminate our white dwarf sample.
Any changes to the completeness arising from the cleaning procedure must be accounted for and applied to the error distribution function from the artificial stars tests.

\subsection{\sharpparam} \label{sec:deepacs_sharpcal}

If an object identified by the \software{DAOPHOT II} program \software{ALLSTAR} is a star, then it should have a \sharpparam\ value near zero.
Values of \sharpparam\ much larger than zero indicate the object is probably a galaxy or unrecognized double, while objects with \sharpparam\ much less than zero are probably cosmic rays or image defects such as bad pixels \citep{daophotii}.
A cut in the \sharpparam\ parameter can thus be used to remove objects that are not stars.
Galaxies in particular are a common contaminant of the white dwarf cooling sequence at very faint magnitudes \citep{2012AJ....143...11K}.

To determine what a reasonable range of \sharpparam\ values is for stars in our sample, we analyse the distribution of \sharpparam\ values for 47 Tuc main-sequence stars as a function of magnitude.
The 47 Tuc white dwarfs should have a distribution in \sharpparam\ similar to that of the 47 Tuc main-sequence stars (at comparable magnitudes), so we use our analysis of the latter to choose the threshold value for our \sharpparam\ cut.
The aim is to make a cut in \sharpparam\ that is generous enough to not reduce the completeness of the white dwarf data but strict enough to remove as many objects that are not stars as possible.

\begin{figure*}
    \centering
    \begin{subfigure}{0.49\textwidth}
    \centering
    \includegraphics{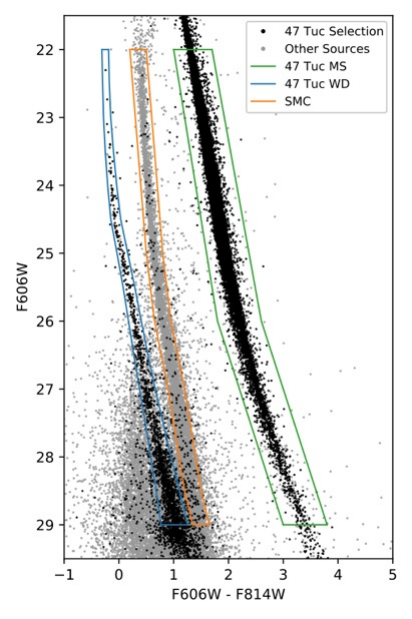}
    \caption{Effect of proper motion cleaning. Not \sharpparam-cleaned.}
    \label{fig:sharpcal_pmcal_cmd_a}
    \end{subfigure}
    \begin{subfigure}{0.49\textwidth}
    \centering
    \includegraphics{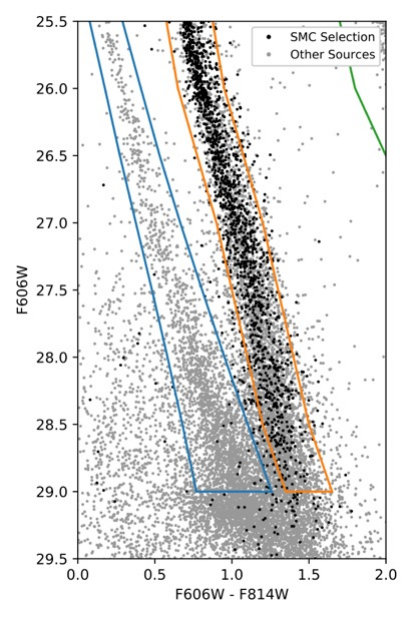}
    \caption{SMC contamination of faint white dwarfs. \sharpparam-cleaned.}
    \label{fig:sharpcal_pmcal_cmd_b}
    \end{subfigure}
    \caption{CMDs showing population boundaries and effects of cleaning procedures.
    The boundaries defining the CMD-selected populations are shown as solid lines. 
    From left to right, these populations are 47 Tuc white dwarfs (blue), SMC stars (orange), and 47 Tuc main-sequence stars (green).
    The left panel shows the effect of proper motion cleaning, while the right panel shows the effect of \sharpparam\ cleaning and the SMC contamination at the faint end of the 47 Tuc white dwarf data space.
    The data point colours indicate different proper motion cuts for the left and right panels.
    \textit{Left:} Data without \sharpparam\ cleaning. Sources selected by the 47 Tuc proper motion cut are shown in black, while rejected sources are shown in grey.
    \textit{Right:} \sharpparam-cleaned data with focus on faint white dwarfs. Sources selected by the SMC proper motion cut are shown in black, while other sources are shown in grey.
    The SMC proper motion cut is used for calibration purposes in \cref{sec:deepacs_pmcal_smc_contam}; its purpose is to select a very pure sample of SMC stars rather than as many SMC members as possible.}
    \label{fig:sharpcal_pmcal_cmd}
\end{figure*}

The 47 Tuc main-sequence stars are selected using a cut in the CMD.
The CMD boundary used to select the 47 Tuc main-sequence stars is shown in \cref{fig:sharpcal_pmcal_cmd} by the green lines (labelled ``47 Tuc MS'').
\Cref{fig:sharpcal_pmcal_cmd} also shows the boundaries used to define CMD-selected populations of 47 Tuc white dwarfs (``47 Tuc WD'', blue lines) and SMC stars (``SMC'', orange lines), which are used later in the analysis in \cref{sec:deepacs_pmcal}.
The CMD boundary for the 47 Tuc white dwarfs is furthermore the same boundary ultimately used to define the data space in the unbinned likelihood analysis. 
In this \lcnamecref{sec:deepacs_sharpcal}, these boundaries of the 47 Tuc white dwarfs and the SMC in the CMD simply serve as visual references to note the locations of these populations.
From left to right across \cref{fig:sharpcal_pmcal_cmd}, the CMD boundaries correspond to 47 Tuc white dwarfs, SMC stars, and 47 Tuc main-sequence stars.
The CMD boundaries of these three populations are the same in both sub-figures of \cref{fig:sharpcal_pmcal_cmd}, though the full span of each bounding region can only be seen in \cref{fig:sharpcal_pmcal_cmd_a}.
All three boundaries span the same range of F606W magnitude values along the y-axis, from 22 to 29, so the calibration analysis in both this \lcnamecref{sec:deepacs_sharpcal} and \cref{sec:deepacs_pmcal} can be done as a function of F606W.

We construct the empirical \sharpparam\ distributions for the CMD-selected main-sequence stars both before and after performing a proper motion cut to select objects within $2.5~\textrm{mas} \, \textrm{yr}^{-1}$ of the mean proper motion of 47 Tuc\footnote{In contrast, the proper motion cut to select SMC stars used to calibrate the proper motion cleaning procedure (discussed in \cref{sec:deepacs_pmcal}) selects objects within $0.75~\textrm{mas} \, \textrm{yr}^{-1}$ of the mean proper motion of the SMC, a much tighter cut in proper motion than what is used for 47 Tuc.}.
This is the same proper motion cut that we use in the final cleaning procedure for our white dwarf data.
The \sharpparam\ distributions for the proper-motion-cleaned main-sequence stars are of most interest to us because the proper motion cleaning produces a more pure sample of 47 Tuc members.
However, comparing the \sharpparam\ distributions both before and after proper motion cleaning is also useful as it gives us information about the typical \sharpparam\ values of the outliers removed by the proper motion cut, especially those sources that are likely not stars.
The reduction in completeness caused by the proper motion cut is analysed in \cref{sec:deepacs_pmcal}. 
While that is an important consideration for our analysis of the white dwarf cooling, a reduction in completeness of the main-sequence sample due to this proper motion cut is not a concern for our analysis of the \sharpparam\ distribution, as it will only affect the amplitude, not the shape, of the distribution.

The effect of the proper motion cleaning on the CMD of the data is shown in \cref{fig:sharpcal_pmcal_cmd_a}, where it can be seen that most of the SMC stars are removed by the cut in proper motion.
While some SMC stars survive the proper motion cut, the SMC is located far enough away from the 47 Tuc main sequence in the CMD that the SMC stars do not contaminate the main-sequence sample selected using the CMD cut.
In \cref{fig:sharpcal_pmcal_cmd_a}, the black points show the sources that were selected as likely 47 Tuc members by the proper motion cut, while the grey points show sources that were rejected by this cut.
Note that the black and grey colour-coding of the data points in \cref{fig:sharpcal_pmcal_cmd_b} has a different meaning than in \cref{fig:sharpcal_pmcal_cmd_a}.
In \cref{fig:sharpcal_pmcal_cmd_b}, the colour-coding of the data points indicates which objects are selected (black) or rejected (grey) by a proper motion cut to select likely SMC members, which is used in \cref{sec:deepacs_pmcal_smc_contam} to analyse the SMC contamination in the 47 Tuc white dwarf data space and is described in detail in that \lcnamecref{sec:deepacs_pmcal_smc_contam}.
It should also be noted that both the CMD and proper motion selections of SMC stars are only used for the purpose of calibrating the proper motion data cleaning procedure in \cref{sec:deepacs_pmcal} and do not need to be complete for this purpose, so these selections prioritize the purity of the SMC sample over the completeness of the sample.
This results in many SMC stars being excluded from the SMC selections, particularly for the proper motion selection of SMC stars shown in \cref{fig:sharpcal_pmcal_cmd_b}.
These SMC selections are not relevant for the \sharpparam\ cleaning procedure discussed in the current section, so the discussion here is kept brief, but more details can be found in \cref{sec:deepacs_pmcal_smc_contam}.

\begin{figure*}
    \centering
    \begin{subfigure}{0.47\textwidth}
        \centering
        \includegraphics{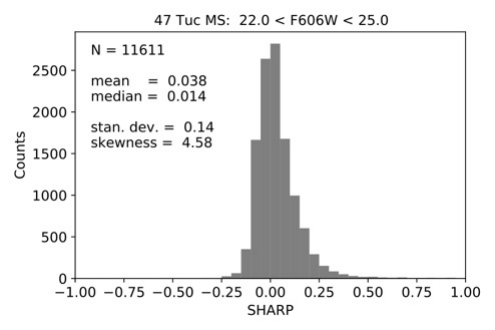}\\
        \includegraphics{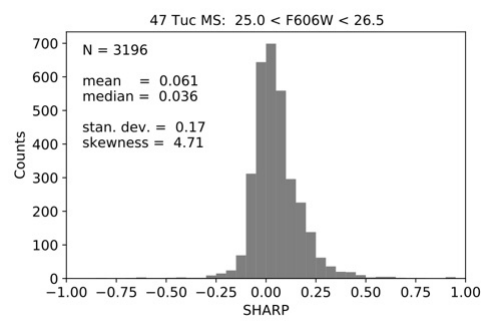}\\
        \includegraphics{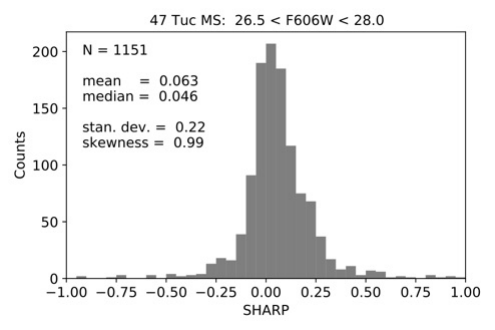}\\
        \includegraphics{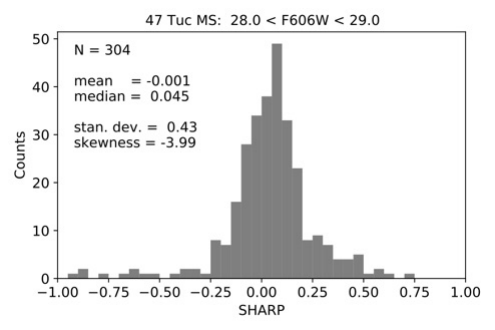}
        \caption{Before proper motion cleaning.}
        \label{fig:sharpcal_dist_a}
    \end{subfigure}
    \hfill
    \begin{subfigure}{0.47\textwidth}
        \centering
        \includegraphics{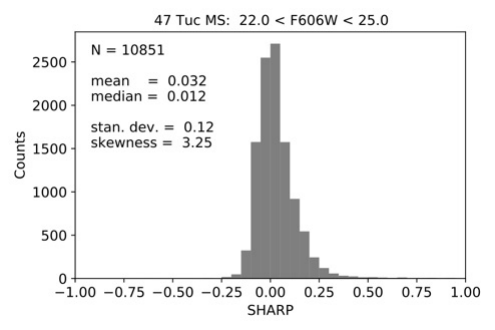}\\
        \includegraphics{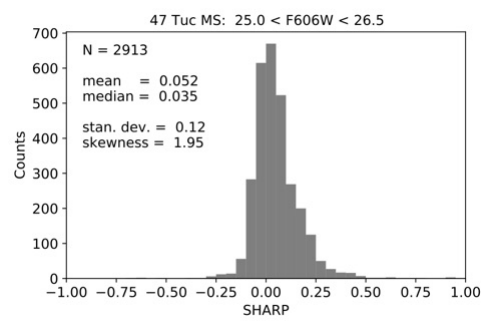}\\
        \includegraphics{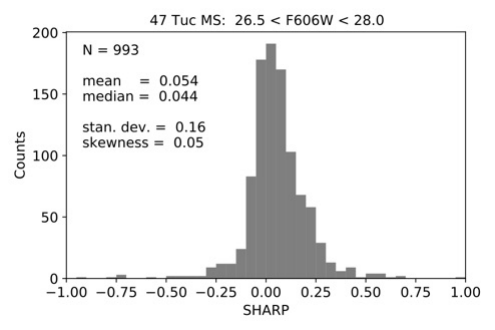}\\
        \includegraphics{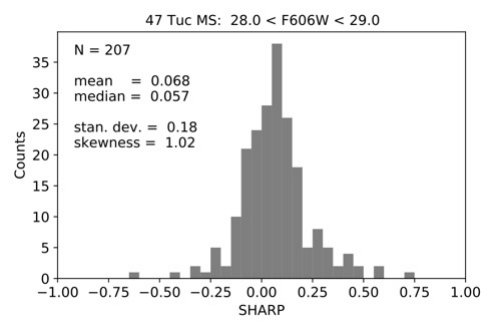}
        \caption{After proper motion cleaning.}
        \label{fig:sharpcal_dist_b}
    \end{subfigure}
    \caption{Distribution of \sharpparam\ for 47 Tuc main-sequence stars by F606W magnitude bin.}
    \label{fig:sharpcal_dist}
\end{figure*}

The empirical number distributions of the \sharpparam\ parameter for the 47 Tuc main-sequence stars sub-divided into various F606W magnitude bins are shown in \cref{fig:sharpcal_dist}. 
The distributions before proper motion cleaning are shown in \cref{fig:sharpcal_dist_a} (left column), while the distributions after proper motion cleaning are shown in \cref{fig:sharpcal_dist_b} (right column).
Each row corresponds to a different magnitude bin.
From top to bottom, the F606W magnitude bins shown in \cref{fig:sharpcal_dist} are $22 - 25$, $25 - 26.5$, $26.5 - 28$, and $28 - 29$.
The ranges for these bins were chosen by first constructing the \sharpparam\ distributions for evenly spaced magnitude bins of 0.5 width and then grouping together adjacent bins for which the morphology of the distributions was similar.
This grouping was done to facilitate visualisation.
All of the \sharpparam\ distributions, plotted as histograms in \cref{fig:sharpcal_dist}, were constructed using the same \sharpparam\ bin width of 0.05.
For each magnitude bin, we also calculated the following sample statistics: mean, median, standard deviation, and skewness.
These sample statistics, along with the total number $N$ of sources in the bin, are reported in \cref{fig:sharpcal_dist} and summarised in \cref{tab:sharpcal_stats}.
Note that the sample mean is denoted as $\overline{\text{\sharpparam}}$, while the median is denoted as $\widetilde{\text{\sharpparam}}$, with a tilde instead of an overline.
The sample standard deviation of the \sharpparam\ values in each magnitude bin is denoted as $\sigma_\text{\sharpparam}$.

\begin{table*}
    \centering
    \caption{Statistics for the \sharpparam\ distributions of CMD-selected 47 Tuc main-sequence stars. For each F606W magnitude bin, the reported quantities are the number of sources ($N$), the mean ($\overline{\text{\sharpparam}}$), the median ($\widetilde{\text{\sharpparam}}$), the standard deviation ($\sigma_\text{\sharpparam}$), and the skewness ($g_1$).}
    \setlength\tabcolsep{6pt}
    \begin{tabular*}{\textwidth}{@{\extracolsep{\fill}}@{\hspace{6pt}} l c r d{3} d{3} d{3} d{2}}
        \toprule
        ~ & F606W
        & \multicolumn{1}{c}{$N$}
        & \multicolumn{1}{c}{\quad $\overline{\text{\sharpparam}}$}
        & \multicolumn{1}{c}{$\widetilde{\text{\sharpparam}}$}
        & \multicolumn{1}{c}{$\sigma_\text{\sharpparam}$}
        & \multicolumn{1}{c}{\quad $g_1$}\\
        \midrule
        \multirow{4}{*}{Before proper motion cleaning}
        & $22.0 - 25.0$ & $11611$ &  0.038 & 0.014 & 0.14 &  4.58\\
        & $25.0 - 26.5$ &  $3196$ &  0.061 & 0.036 & 0.17 &  4.71\\
        & $26.5 - 28.0$ &  $1151$ &  0.063 & 0.046 & 0.22 &  0.99\\
        & $28.0 - 29.0$ &   $304$ & -0.001 & 0.045 & 0.43 & -3.99\\
        \midrule
        \multirow{4}{*}{After proper motion cleaning}
        & $22.0 - 25.0$ & $10851$ &  0.032 & 0.012 & 0.12 &  3.25\\
        & $25.0 - 26.5$ &  $2913$ &  0.052 & 0.035 & 0.12 &  1.95\\
        & $26.5 - 28.0$ &   $993$ &  0.054 & 0.044 & 0.16 &  0.05\\
        & $28.0 - 29.0$ &   $207$ &  0.068 & 0.057 & 0.18 &  1.02\\
        \bottomrule
    \end{tabular*}
    \label{tab:sharpcal_stats}
\end{table*}

For the sample skewness statistic, we use the Fisher-Pearson coefficient of skewness
\begin{equation}
    g_1 = \frac{m_3}{m_2^{3/2}},
\end{equation}
which is calculated from the biased sample second and third central moments, respectively $m_2$ and $m_3$, where
\begin{equation}
    m_n = \frac{1}{N} \sum_{i=1}^N \left( \text{\sharpparam}_i - \overline{\text{\sharpparam}} \right)^n
\end{equation}
is the biased sample $n$th central moment.
The sample sizes are large enough that using the adjusted Fisher-Pearson standardized moment coefficient, $G_1 = g_1 \sqrt{N\left(N-1\right)}/\left(N-2\right)$, instead of $g_1$ to correct for bias makes negligible difference to the results. 
To the level of significance reported, the only difference is in the faintest bin, where $G_1 = -4.01$ (instead of $g_1 = -3.99$) before proper motion cleaning and $G_1 = 1.03$ (instead of $g_1 = 1.02$) after proper motion cleaning.

The mean and median are both measures of central tendency, with the median being more robust to outliers.
Both the sample mean and sample median are close to zero for all magnitude bins, as expected.
This applies for the \sharpparam\ distributions both before and after proper motion cleaning.
For the standard deviation, which is a measure of the spread of \sharpparam\ values, we see that proper motion cleaning makes more of a difference.
Before proper motion cleaning, $\sigma_\text{\sharpparam}$ increases more dramatically with magnitude; while after proper motion cleaning, the $\sigma_\text{\sharpparam}$ values remain more similar even down to the faintest magnitude bin.
$\sigma_\text{\sharpparam}$ is also overall smaller after proper motion cleaning, though the value is similar before and after proper motion cleaning for the brightest stars.
After proper motion cleaning, we see that a \sharpparam\ cut of $\vert \text{\sharpparam} \vert < 0.5$ corresponds to a cut of $\gtrsim 3 \, \sigma_\text{\sharpparam}$ for all magnitude bins.

Even before proper motion cleaning, we find that $\sigma_\text{\sharpparam}$ for the two magnitude bins containing the brightest sources ($22.0 < \text{F606W} < 25.0$ and $25.0 < \text{F606W} < 26.5$) are small enough that $3 \, \sigma_\text{\sharpparam} \lesssim 0.5$.
The two magnitude bins containing the faintest sources ($26.5 < \text{F606W} < 28.0$ and $28.0 < \text{F606W} < 29.0$) have larger estimated values of the standard deviation, but they also have much larger fractions of outliers that are likely not actually stars (particularly the faintest bin, $28.0 < \text{F606W} < 29.0$, which has the largest estimated standard deviation).
These outliers can be seen in the tails of the distributions in \cref{fig:sharpcal_dist_a}, and comparison of those distributions with the distributions for the same magnitude bins in \cref{fig:sharpcal_dist_b} shows that most of these outliers are removed by the proper motion cut.
The sample standard deviation is not very robust to outliers, so for these bins in particular, and before proper motion cleaning overall, the outliers lead to an overestimate of the standard deviation of the underlying \sharpparam\ distribution for sources that are actually single stars.
The values of $\sigma_\text{\sharpparam}$ after proper motion cleaning should be a better estimate, and we indeed find these values to be smaller than the values of $\sigma_\text{\sharpparam}$ before proper motion cleaning.

While the \sharpparam\ values are approximately normally-distributed, it can be seen from \cref{fig:sharpcal_dist} that the \sharpparam\ distributions have some asymmetry.
The skewness statistic is a measure of this asymmetry.
Before proper motion cleaning, the \sharpparam\ distributions for the brightest two magnitude bins have a similar value of skewness, both being positively skewed with a longer tail for positive \sharpparam\ values.
This positive skewness may be in part due to the presence of unresolved binaries, which would have large positive \sharpparam\ values \citep{daophotii} and would not be removed by proper motion cleaning.
The negative skewness for the faintest bin ($28.0 < \text{F606W} < 29.0$) before proper motion cleaning, on the other hand, can be attributed to the long tail of likely non-stellar outliers with large negative \sharpparam\ values.
After proper motion cleaning, we see that the skewness in the faintest bin becomes positive, like the skewness in the other magnitude bins, and the absolute value of the skewness decreases in all of the bins.

For all magnitudes, we find that most of the objects are contained within the range $-0.5 < \text{\sharpparam} < 0.5$.
The small number of objects outside of this \sharpparam\ range are outliers of the \sharpparam\ distributions that are unlikely to be stars. 
We thus choose a cut in the \sharpparam\ parameter of $\lvert\text{\sharpparam}\rvert < 0.5$ for the cleaning procedure of our main white dwarf data.
The result of this \sharpparam\ cleaning is shown in \cref{fig:sharpcal_pmcal_cmd_b}, which focuses on faint magnitudes where the SMC sequence begins to overlap with the 47 Tuc white dwarf cooling sequence.
The improvement in data quality achieved by \sharpparam\ cleaning can be seen by comparing \cref{fig:sharpcal_pmcal_cmd_b} to \cref{fig:sharpcal_pmcal_cmd_a}, the latter of which shows the data before any \sharpparam\ cleaning.
The black and grey colour-coding of the data points in the two sub-figures sorts the data based on different proper motion cuts in each sub-figure, but these colours do not indicate anything about \sharpparam\ cleaning in either case. 
Neither the proper-motion-selected nor proper-motion-rejected data points shown in \cref{fig:sharpcal_pmcal_cmd_a} have been \sharpparam-cleaned, while both the proper-motion-selected and proper-motion-rejected data points shown in \cref{fig:sharpcal_pmcal_cmd_b} have been \sharpparam-cleaned so that only points for which $\lvert\text{\sharpparam}\rvert < 0.5$ are shown in \cref{fig:sharpcal_pmcal_cmd_b}.

\subsection{Proper Motion} \label{sec:deepacs_pmcal}

A cut in proper motion allows us to remove most of the SMC stars (and other contaminants like galaxies) from our white dwarf sample.
This is particularly useful for cleaning our white dwarf sample at the faint end of the cooling sequence where it intersects with the SMC sequence in the CMD (see \cref{fig:sharpcal_pmcal_cmd}).
However, the 47 Tuc and SMC populations overlap in the tails of their proper motion distributions, so the SMC contaminants cannot be completely removed with a proper motion cut, at least not for the faintest stars of interest to us where the larger proper motion errors lead to more diffuse proper motion distributions for both the 47 Tuc and SMC populations. 
Furthermore, a cut in proper motion reduces the completeness of our sample.
We quantify both of these effects in this \lcnamecref{sec:deepacs_pmcal}.

We first quantify the residual SMC contamination using both CMD-selected and proper-motion-selected SMC stars. 
Then we quantify the reduction in completeness using CMD-selected 47 Tuc main-sequence stars.
For both of these procedures, we first clean our data by applying the \sharpparam\ parameter cut determined in \cref{sec:deepacs_sharpcal}, i.e. $\lvert\text{\sharpparam}\rvert < 0.5$, to the initial catalogue from the deep HST ACS/WFC observations.
Based on the analysis of \cref{sec:deepacs_sharpcal}, this cut does not reduce the completeness of our 47 Tuc sample, but it removes non-stellar contaminants and facilitates identification of the CMD populations, particularly at the faint end of the 47 Tuc white dwarf and SMC sequences.

\subsubsection{Proper Motion Distribution Model}  \label{sec:deepacs_pmcal_pm_dist}

To inform our choice of proper motion cuts in our analysis of the proper motion cleaning procedure, we want to know the mean proper motion of the SMC relative to the mean proper motion of 47 Tuc. We also want to know the spread of proper motion values for each of these two populations, which is quantified by the standard deviation for a population with normally-distributed proper motions.
Let $\mu_\alpha$ and $\mu_\delta$ be the components of the tangent plane projection of the proper motion vector, where $\mu_\alpha$ is the component in the direction of increasing right ascension and $\mu_\delta$ is the component in the direction of increasing declination. 
While the mean proper motion of 47 Tuc has already been subtracted from our proper motion data, this mean proper motion is still included in the model discussed in this \lcnamecref{sec:deepacs_pmcal_pm_dist} to make the dependence on the mean motion of 47 Tuc explicit.

We model the distribution of proper motions for our data as a three-{\allowbreak}component Gaussian mixture model.
In this model, the joint probability density function of $\mu_\alpha$ and $\mu_\delta$ is taken to be the linear superposition of three bivariate normal distributions, each with its own mean $(\bar{\mu}_{\alpha,i}, \bar{\mu}_{\delta,i})$, standard deviation $\sigma_i$, and amplitude $A_i$, where $i$ is an index labelling the constituent distributions.
One of these Gaussian components accounts for the 47 Tuc population ($i=1$), another one accounts for the SMC population ($i=2$), and the final one accounts for outliers and background contaminants like field stars ($i=3$).
Written explicitly, the joint probability density function of $\mu_\alpha$ and $\mu_\delta$ is
\begin{align}
\begin{split}
    &f_{\mu_\alpha,\mu_\delta}\left(\mu_\alpha,\mu_\delta; \, \theta \right)\\
    &\qquad = \sum_{i=1}^3 A_i \, f_{\mu_\alpha,\mu_\delta,i}\left(\mu_\alpha,\mu_\delta; \, \bar{\mu}_{\alpha,i}, \bar{\mu}_{\delta,i}, \sigma_i \right),
\end{split}
\label{eq:pm_3comp_GMM_pdf}
\end{align}
where $\theta$ denotes the full set of parameters that characterise the distribution and the probability density distribution of a single population labelled with index $i$ is
\begin{align}
\begin{split}
    &f_{\mu_\alpha,\mu_\delta,i}\left(\mu_\alpha,\mu_\delta; \, \bar{\mu}_{\alpha,i}, \bar{\mu}_{\delta,i}, \sigma_i \right)\\
    &\qquad = \frac{1}{2 \pi \sigma_i} \exp\left[-\frac{\left(\mu_\alpha - \bar{\mu}_{\alpha,i}\right)^2}{2\sigma_i^2} -\frac{\left(\mu_\delta - \bar{\mu}_{\delta,i}\right)^2}{2\sigma_i^2}\right].
\end{split}
\label{eq:pm_dist_GMM_single_pop}
\end{align}
It has been assumed that the two proper motion components $\mu_\alpha$ and $\mu_\delta$ are uncorrelated and have the same standard deviation for a given population.
In general, these assumptions need not be true and could be relaxed in the model, but the current model given by \cref{eq:pm_3comp_GMM_pdf} is sufficient for our purposes.
The model is also further simplified by eliminating the dependence of $f_{\mu_\alpha,\mu_\delta}$ on a few parameters in \cref{eq:pm_3comp_GMM_pdf} as follows.

The proper motion data are already given relative to the mean proper motion of 47 Tuc, so we keep the mean proper motion of 47 Tuc fixed in this model with the value $(\bar{\mu}_{\alpha,1}, \bar{\mu}_{\delta,1}) = (0, 0)$.
As the total probability density function must be normalised to unity and each of the constituent Gaussian distributions is normalised to unity, the sum of amplitudes must be equal to unity, i.e. $\sum_i A_i = 1$.
This relation eliminates dependence of the total probability density function on one of the amplitudes.
We choose to eliminate $A_3$ by setting $A_3 = 1 - A_1 - A_2$.
All together, this reduces the number of parameters that $f_{\mu_\alpha,\mu_\delta}$ depends on in \cref{eq:pm_3comp_GMM_pdf} from 12 to 9. 
The set of remaining parameters is 
\begin{equation}
    \theta = \left\lbrace A_1, \sigma_1, A_2, \sigma_2, \bar{\mu}_{\alpha,2}, \bar{\mu}_{\delta,2}, \sigma_3, \bar{\mu}_{\alpha,3}, \bar{\mu}_{\delta,3} \right\rbrace.
\end{equation}

The best-fitting values of the remaining 9 parameters, including the mean proper motion components of the SMC, are then determined using the maximum likelihood estimate.
Let $d = \left\lbrace d_j \right\rbrace$ be the set of observed proper motion data points, where $d_j = \left(\mu_{\alpha j}, \ \mu_{\delta j}\right)$ is a single data point in proper motion space and $j$ is an index that labels the data points.
The likelihood $\mathcal{L}\left( \theta \right)$ is the probability (density) of the observed data given the parameters and model,
\begin{align}
    \mathcal{L}\left(\theta\right)
    &= p\left(d | \theta \right)\\
    &= \prod_j p\left(d_j | \theta \right)\\
    &= \prod_j f_{\mu_\alpha,\mu_\delta}\left(\mu_{\alpha j},\mu_{\delta j}; \, \theta\right).
\end{align}
The natural logarithm of the likelihood for our model of the proper motion distribution is thus
\begin{equation}
    \ln \mathcal{L}\left( \theta \right) = \sum_j \ln \, f_{\mu_\alpha,\mu_\delta}\left(\mu_{\alpha j},\mu_{\delta j}; \, \theta\right).
    \label{eq:pm_dist_lnL}
\end{equation}
The maximum likelihood estimate $\hat{\theta}$ of the model parameters is the set of parameter values that maximises $\mathcal{L}\left(\theta\right)$, i.e. that maximises the probability of the observed data.
In practice, it is more computationally feasible to minimise the negative log-likelihood, $-\ln\mathcal{L}\left(\theta\right)$, and doing so is equivalent to maximising $\mathcal{L}\left(\theta\right)$.
We minimise the negative of \cref{eq:pm_dist_lnL} numerically to get $\hat{\theta}$, which is the set of best-fitting parameters for our model.

\begin{table*}
    \centering
    \caption{Results of fitting proper motion distribution by F606W magnitude bin. The subscript indices of the parameters denote which population that parameter describes in a three-component Gaussian mixture model: 1 denotes 47 Tuc, 2 denotes the SMC, and 3 denotes the background. The average proper motion coordinates of the SMC across all magnitude bins are $\bar{\mu}_{\alpha,2} = 4.76$ and $\bar{\mu}_{\delta,2} = 1.59$.}
    \setlength\tabcolsep{6pt}
    \begin{tabular*}{\textwidth}{@{\extracolsep{\fill}}@{\hspace{6pt}} c  d{1.2} d{1.2}  d{1.2} d{1.2} d{1.2} d{1.2}  d{1.2} d{1.2} d{2.2} d{2.2}}
        \toprule
        \heading{F606W} & \heading{$A_1$} & \heading{$\sigma_1$} & \heading{$A_2$} & \heading{$\sigma_2$} & \heading{$\bar{\mu}_{\alpha,2}$} & \heading{$\bar{\mu}_{\delta,2}$} & \heading{$A_3$} & \heading{$\sigma_3$} & \heading{~~$\bar{\mu}_{\alpha,3}$} & \heading{~~$\bar{\mu}_{\delta,3}$}\\
        \midrule
        $22.0-22.5$  &   0.82 & 0.62  &   0.09 & 0.44 & 4.75 & 1.53  &   0.10 & 3.34 & -1.07 &  0.39\\
        $22.5-23.0$  &   0.77 & 0.60  &   0.10 & 0.47 & 4.74 & 1.49  &   0.13 & 2.76 & -0.76 &  0.00\\
        $23.0-23.5$  &   0.78 & 0.60  &   0.10 & 0.43 & 4.67 & 1.44  &   0.12 & 3.04 & -0.82 & -0.05\\
        $23.5-24.0$  &   0.75 & 0.61  &   0.12 & 0.53 & 4.72 & 1.52  &   0.13 & 2.51 & -0.93 & -0.15\\
        $24.0-24.5$  &   0.75 & 0.62  &   0.14 & 0.48 & 4.75 & 1.48  &   0.11 & 2.99 & -0.57 &  0.06\\
        $24.5-25.0$  &   0.70 & 0.64  &   0.17 & 0.52 & 4.71 & 1.51  &   0.13 & 3.27 & -0.29 &  0.26\\
        $25.0-25.5$  &   0.63 & 0.64  &   0.22 & 0.56 & 4.73 & 1.56  &   0.15 & 3.55 &  0.83 &  0.73\\
        $25.5-26.0$  &   0.56 & 0.69  &   0.28 & 0.62 & 4.73 & 1.61  &   0.16 & 4.28 &  0.65 &  0.40\\
        $26.0-26.5$  &   0.48 & 0.74  &   0.37 & 0.81 & 4.74 & 1.69  &   0.15 & 5.82 &  1.02 &  0.68\\
        $26.5-27.0$  &   0.35 & 0.82  &   0.43 & 0.93 & 4.75 & 1.67  &   0.22 & 6.27 &  2.06 &  1.15\\
        $27.0-27.5$  &   0.27 & 0.90  &   0.49 & 1.15 & 4.86 & 1.61  &   0.24 & 6.83 &  1.18 &  0.40\\
        $27.5-28.0$  &   0.21 & 1.08  &   0.49 & 1.37 & 4.84 & 1.78  &   0.30 & 7.94 &  1.33 &  0.49\\
        $28.0-28.5$  &   0.20 & 1.36  &   0.44 & 1.55 & 4.71 & 1.67  &   0.35 & 8.32 &  1.44 &  0.57\\
        $28.5-29.0$  &   0.22 & 1.63  &   0.35 & 1.79 & 4.88 & 1.68  &   0.43 & 8.38 &  0.92 &  0.46\\
        \bottomrule
    \end{tabular*}
    \label{tab:47tuc_deepacs_pm_fit}
\end{table*}

The three-component Gaussian mixture model was fit to the proper motions for different F606W magnitude bins spanning the range $22 - 29$ in increments of $0.5$. The maximum likelihood estimates of the distribution parameters for each of the magnitude bins are given in \cref{tab:47tuc_deepacs_pm_fit}.
The standard deviations of both the 47 Tuc population and the SMC population increase with magnitude due to the increasing error in the proper motion measurements with magnitude.
Note that the errors can be taken to be Gaussian-distributed and the convolution of two Gaussians is another Gaussian, so the proper motion errors are naturally accounted for in the Gaussian mixture model as an adjustment to the standard deviations of the populations.
These increasing proper motion errors do not have a notable effect on the location of the SMC in proper motion space, which was found to be similar across all of the magnitude bins.
The coordinates of the SMC in proper motion space are taken to be the average of the coordinates found in each bin.
These coordinates are $\left(\bar{\mu}_{\alpha,2}, \bar{\mu}_{\delta,2}\right) = \left(4.76, 1.59\right)$.

Though the best-fitting parameters are determined using the joint distribution of $\mu_\alpha$ and $\mu_\delta$, it is also instructive to see how the total proper motion is distributed for a particular population in polar coordinates after marginalising over the polar angle.
This marginal distribution is derived below, and some of its important features are discussed.

Consider a population labelled by index $i$ with proper motion distribution given by \cref{eq:pm_dist_GMM_single_pop}.
Define the polar coordinate variables $\mu$ and $\phi$ relative to the mean proper motion of this population, such that
\begin{alignat}{3}
    &\mu_\alpha &&- \bar{\mu}_{\alpha,i} &&= \mu \cos\phi,\\
    &\mu_\delta &&- \bar{\mu}_{\delta,i} &&= \mu \sin\phi.
\end{alignat}
The determinant of the Jacobian matrix $\mat{J}(\mu,\phi)$ for the transformation from the Cartesian to polar coordinates is $\mathrm{det}\left\lvert \mat{J}(\mu,\phi) \right\rvert = \mu$.

Accounting for the relevant factor of $\mathrm{det}\left\lvert \mat{J}(\mu,\phi) \right\rvert$, the joint probability density function of $\mu$ and $\phi$ is thus given by the relation
\begin{align}
\begin{split}
    &f_{\mu,\phi,i}\left(\mu, \phi; \, \sigma_i\right)\\
    &\ \ = \mu \ f_{\mu_\alpha,\mu_\delta,i}\left(\bar{\mu}_{\alpha,i}+\mu\cos\phi, \bar{\mu}_{\delta,i}+\mu\sin\phi; \, \bar{\mu}_{\alpha,i}, \bar{\mu}_{\delta,i}, \sigma_i\right)
\end{split}\label{eq:pm_dist_fmuphi1}\\
    &\ \ = \frac{\mu}{2\pi\sigma_i^2} \exp\left(-\frac{\mu^2}{2\sigma_i^2}\right).
\end{align}
Marginalising $f_{\mu,\phi,i}\left(\mu, \phi; \, \sigma_i\right)$ over $\phi$ then gives the probability density function of $\mu$,
\begin{align}
    f_{\mu,i}\left(\mu; \, \sigma_i\right) 
    &= \int_{0}^{2\pi} \mathrm{d}\phi \ f_{\mu,\phi,i}\left(\mu, \phi; \, \sigma_i\right)\\
    &= \frac{\mu}{\sigma_i^2} \exp\left(-\frac{\mu^2}{2\sigma_i^2}\right).\label{eq:pm_dist_fmui}
\end{align}
This function goes to zero in the limit that $\mu$ goes to zero, i.e. $\lim_{\mu \rightarrow 0} f_{\mu,i}(\mu; \, \sigma_i) = 0$.
Also note that the probability density of $\mu$ is maximized when $\mu = \sigma_i$; this is in contrast to the joint probability density of $\mu_\alpha$ and $\mu_\delta$, which is maximised at the coordinates $(\bar{\mu}_{\alpha,i}, \bar{\mu}_{\delta,i})$ and thus $\mu = 0$.

If proper motion values are given relative to the mean motion of 47 Tuc, then the probability density distribution of $\mu$ values for the 47 Tuc population goes to zero as $\mu \rightarrow 0$.
This is straight-forwardly given by \cref{eq:pm_dist_fmui} with $i=1$, as the mean proper motion of the 47 Tuc population coincides with the origin of the polar coordinate system in proper motion space.
For the SMC population, the distribution of the total proper motion relative to the mean proper motion of 47 Tuc is more complicated, as the mean proper motion of the SMC does not coincide with the origin of the polar coordinate system.
However, the factor of $\mu$ that appears in \cref{eq:pm_dist_fmuphi1}, which is the determinant of the Jacobian in transforming the proper motion distribution function from Cartesian coordinates to polar coordinates, also appears in the analogous distribution for the SMC when $\mu$ is defined with respect to $\left(\bar{\mu}_{\alpha,1}, \ \bar{\mu}_{\delta,1} \right)$ instead of $\left(\bar{\mu}_{\alpha,2}, \ \bar{\mu}_{\delta,2} \right)$, and this factor likewise results in the distribution of $\mu$ for SMC stars going to zero as $\mu \rightarrow 0$.
The SMC population is also far enough away from the 47 Tuc population in proper motion space that very few SMC stars are expected to be found at the centre of the 47 Tuc distribution in proper motion space in the first place.

\subsubsection{SMC Contamination}  \label{sec:deepacs_pmcal_smc_contam}

While the \sharpparam\ cleaning makes the 47 Tuc white dwarf cooling sequence and the SMC sequence appear more distinct in the CMD (see \cref{fig:sharpcal_pmcal_cmd}), these sequences still intersect at faint magnitudes (for F606W greater than about 27).
The proper motion cut to select likely 47 Tuc members, which will be referred to as simply the ``47 Tuc proper motion cut'', also does not remove all of the SMC stars that overlap with the faint white dwarfs in the cooling sequence. 
This is illustrated by \cref{fig:sharpcal_pmcal_cmd_a}, where it can be seen that some of the objects selected by the 47 Tuc proper motion cut (black points) lie along the SMC sequence (mostly composed of grey points).
Though \cref{fig:sharpcal_pmcal_cmd_a} has not been \sharpparam-cleaned, most of the black points along the SMC sequence in \cref{fig:sharpcal_pmcal_cmd_a} persist after \sharpparam\ cleaning, so the problem of SMC contamination remains.
We want to quantify the number of SMC stars that are expected to be in the final fully-cleaned white dwarf sample that we use for the unbinned likelihood analysis.
The number of SMC contaminants in the white dwarf data space is in general a function of magnitude, so we determine this number for F606W magnitude bins, using bins of 0.5 width spanning the range $22-29$.

To quantify the number of SMC stars expected to contaminate our final white dwarf sample, we define cuts in both the CMD and proper motion space that each independently select stars that are very likely to be SMC members.
The ratio of the number of SMC stars in the 47 Tuc white dwarf CMD region vs the SMC CMD region should be the same regardless of what proper motion cut is used (as long as some SMC stars survive the proper motion cut). Thus, the number of SMC contaminants in the white dwarf CMD region after the 47 Tuc proper motion cut can be estimated by calculating this ratio using a very pure sample of SMC stars selected using a proper motion cut and multiplying this ratio by the number of stars in the SMC CMD region after the 47 Tuc proper motion cut.

Let $\Npscw$ be the number of stars that survive both the tight proper motion cut to select SMC stars and the CMD cut to select white dwarfs in 47 Tuc.
Let $\Npscs$ be the number of stars that survive the same SMC proper motion cut used to get $\Npscw$ and that also survive the CMD cut to select SMC stars.
Finally, let $\Nptcs$ be the number of stars that survive the 47 Tuc proper motion cut and also survive the same CMD cut to select SMC stars used to get $\Npscs$.
Then the number of SMC stars expected to survive both the 47 Tuc proper motion cut and the 47 Tuc white dwarf CMD cut is
\begin{equation}
    \Ncontam = \frac{\Npscw \ \Nptcs}{\Npscs}.
    \label{eq:Ncontam}
\end{equation}
This is the expected number of SMC contaminants in the proper-motion-cleaned white dwarf sample that we ultimately use in the unbinned likelihood analysis.
A formal treatment of the derivation of \cref{eq:Ncontam} is given in \cref{sec:appendix_smc_contamination}.

The goal of both the SMC proper motion cut and the SMC CMD cut is to get a pure sample of SMC stars, and the boundaries of these cuts are chosen with this goal in mind. However, it is still possible that some 47 Tuc stars could survive these cuts, particularly the SMC CMD cut where the SMC sequence and 47 Tuc white dwarf cooling sequence begin to overlap.
These misclassified 47 Tuc stars would cause our count for the corresponding number used in the calculation of $\Ncontam$ to be too large.
This is most likely to be an issue in determining the number of stars that survive one 47 Tuc cut and one SMC cut, i.e. $\Npscw$ or $\Nptcs$, and in particular $\Nptcs$.
It is least likely to be an issue in determining the number of stars that survive both SMC cuts, i.e. $\Npscs$, as these stars should be the purest sample of SMC stars.
Note that both $\Npscw$ and $\Nptcs$ appear in the numerator on the right-hand side of \cref{eq:Ncontam}, while $\Npscs$ appears in the denominator.
$\Ncontam$ is thus more properly an upper limit on the number of SMC stars that contaminate the 47 Tuc white dwarf data space after the 47 Tuc proper motion cut.
As the boundaries for the SMC cuts are specifically chosen to reduce the risk of misclassifying 47 Tuc stars, the true number of SMC contaminants in the proper-motion-cleaned white dwarf data space should be close to this upper limit.
Furthermore, if this upper limit is found to be negligibly small compared to the total size of the white dwarf sample, then that is sufficient information to deem the possibility of SMC contamination in the white dwarf data space to be of no further concern.

The boundaries defining the CMD-selected SMC sample are shown in \cref{fig:sharpcal_pmcal_cmd} as the middle boundary region (orange lines).
The CMD boundaries for the SMC population select predominantly the red side of the SMC sequence at faint magnitudes in order to avoid including 47 Tuc white dwarfs in the CMD-selected SMC population. 
The priority here is to select a pure population of SMC stars, even if it results in the exclusion of some SMC members.
This SMC sample does not need to be complete for our analysis.
The boundary region defining the 47 Tuc white dwarf CMD selection is also shown in \cref{fig:sharpcal_pmcal_cmd}, as the left-most boundary region (blue lines).
This CMD boundary region for 47 Tuc white dwarfs is the same as the white dwarf data space that will be used in the unbinned likelihood analysis.
In \cref{fig:sharpcal_pmcal_cmd_b}, objects in the \sharpparam-cleaned data that survive the SMC proper motion cut are shown as black points, while the other objects in that data are shown as grey points.

\begin{figure}
    \centering
    \begin{subfigure}{\columnwidth}
    \centering
    \includegraphics[clip,trim={0 0.73cm 0 1.29cm}]{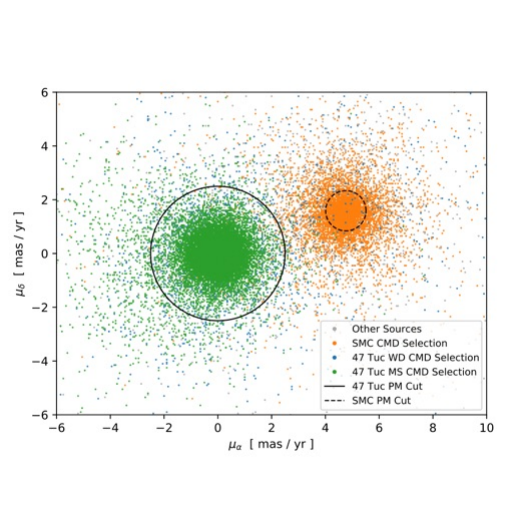}
    \caption{Components of proper motion vector in tangent plane.}
    \label{fig:pmcal_cmd_pops_pm_a}
    \end{subfigure}\\[1ex]
    \begin{subfigure}{\columnwidth}
    \centering
    \includegraphics{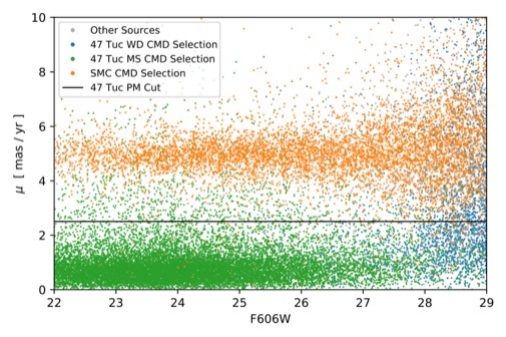}
    \caption{Total proper motions as a function of F606W magnitude.}
    \label{fig:pmcal_cmd_pops_pm_b}
    \end{subfigure}
    \caption{Proper motions of \sharpparam-cleaned data relative to mean motion of 47 Tuc.
    The CMD-selected population to which each source belongs is indicated by colour: 47 Tuc white dwarfs (blue), 47 Tuc main-sequence stars (green), SMC stars (orange), and other sources with $22 < \text{F606W} < 29$ but not in a CMD boundary region (grey).
    The boundary of the proper motion cut to select 47 Tuc members is shown as a solid black curve indicating a circle of radius $2.5~\textrm{mas}~\textrm{yr}^{-1}$ centred on the mean 47 Tuc proper motion coordinates (0, 0). The boundary of the tight SMC proper motion cut used in the proper motion calibration procedure is shown in sub-figure (a) as a dashed black curve of radius $0.75~\textrm{mas}~\textrm{yr}^{-1}$ centred on the mean motion of the SMC at (4.76, 1.59).}
    \label{fig:pmcal_cmd_pops_pm}
\end{figure}

The proper motions for all sources in our \sharpparam-cleaned data in a frame relative to the mean motion of 47 Tuc are shown in \cref{fig:pmcal_cmd_pops_pm}.
For objects that lie within one of the three CMD boundary regions shown in \cref{fig:sharpcal_pmcal_cmd}, the CMD-selected population to which each object belongs is indicated by colour.
These CMD-selected populations are 47 Tuc white dwarfs (blue), 47 Tuc main-sequence stars (green), and SMC stars (orange).
Objects that do not correspond to any of these three CMD-selected populations are shown in grey.
The boundary of the 47 Tuc proper motion cut is shown in both \cref{fig:pmcal_cmd_pops_pm_a} and \cref{fig:pmcal_cmd_pops_pm_b} as a solid black curve.
The boundary of the tight SMC proper motion cut is shown in \cref{fig:pmcal_cmd_pops_pm_a} as a dashed black curve.

Note that the 47 Tuc white dwarfs and main-sequence stars largely overlap in the proper motion plots of \cref{fig:pmcal_cmd_pops_pm}, as expected since they belong to the same dynamical population.
Most of the white dwarfs are obscured in \cref{fig:pmcal_cmd_pops_pm_a} and at bright magnitudes in \cref{fig:pmcal_cmd_pops_pm_b} simply because the main-sequence stars have been plotted on top of them.
However, some of the white dwarfs of most interest to us are still visible, especially those at faint magnitudes in \cref{fig:pmcal_cmd_pops_pm_b} where the number density of the main-sequence stars is much lower than at brighter magnitudes.
In \cref{fig:pmcal_cmd_pops_pm_a}, it can also be seen that some of the objects selected by the 47 Tuc white dwarf CMD cut fall within the boundary of the SMC proper motion cut.
These objects correspond to the black points in \cref{fig:sharpcal_pmcal_cmd_b} that lie within the white dwarf CMD boundary.
As the SMC proper motion cut is very tight about the mean motion of the SMC and far from the bulk of the 47 Tuc proper motion distribution, these objects are highly likely to be SMC stars that lie within the 47 Tuc white dwarf CMD boundary region.

\cref{fig:pmcal_cmd_pops_pm_a} shows the two-dimensional distribution of the components of proper motion in the directions of right ascension and declination, respectively $\mu_\alpha$ and $\mu_\delta$.
Note that a factor of $\cos\delta$ is included in the definition of $\mu_\alpha$ (i.e. $\mu_\alpha = \dot{\alpha} \cos\delta$ and $\mu_\delta = \dot{\delta}$, where the overdot denotes a derivative with respect to time), making $(\mu_\alpha, \mu_\delta)$ the tangent plane projection of the proper motion vector.
Two distinct populations are clearly visible in \cref{fig:pmcal_cmd_pops_pm_a}. 
The population on the left and centred at $(0, 0)$ corresponds to 47 Tuc, while the population on the right corresponds to the SMC.

The 47 Tuc proper motion cut selects objects with
\begin{equation}
    \sqrt{\left(\mu_\alpha - \bar{\mu}_{\alpha,t}\right)^2 + \left(\mu_\delta - \bar{\mu}_{\delta,t}\right)^2} < 2.5~\textrm{mas}~\textrm{yr}^{-1},
\end{equation}
where $(\bar{\mu}_{\alpha,t}, \bar{\mu}_{\delta,t})$ is the mean proper motion of 47 Tuc, which has the value $(\bar{\mu}_{\alpha,t}, \bar{\mu}_{\delta,t}) = (0, 0)$ in the reference frame of \cref{fig:pmcal_cmd_pops_pm}.
This corresponds to selecting the points in \cref{fig:pmcal_cmd_pops_pm_a} that lie within the circle of radius $2.5$ centred on $(0, 0)$, which is shown as a solid black line.

The SMC proper motion cut selects objects with
\begin{equation}
    \sqrt{\left(\mu_\alpha - \bar{\mu}_{\alpha,s}\right)^2 + \left(\mu_\delta - \bar{\mu}_{\delta,s}\right)^2} < 0.75~\textrm{mas}~\textrm{yr}^{-1},
\end{equation}
where $(\bar{\mu}_{\alpha,s}, \bar{\mu}_{\delta,s})$ is the mean proper motion of the SMC, which we take to be $(\bar{\mu}_{\alpha,s}, \bar{\mu}_{\delta,s}) = (4.76, 1.59)$ in the reference frame of \cref{fig:pmcal_cmd_pops_pm}.
This corresponds to selecting the points in \cref{fig:pmcal_cmd_pops_pm_a} that lie within the circle of radius $0.75$ centred on $(4.76, 1.59)$, which is shown as a dashed black line.

\cref{fig:pmcal_cmd_pops_pm_b} shows the total proper motion in the tangent plane as a function of F606W magnitude. 
This total proper motion is defined in terms of $\mu_\alpha$ and $\mu_\delta$ as
\begin{equation}
    \mu = \sqrt{\mu_\alpha^2 + \mu_\delta^2}.\label{eq:total_pm}
\end{equation}
Since the mean proper motion of 47 Tuc is located at proper motion coordinates $(0, 0)$, this is the total proper motion relative to the 47 Tuc mean.
As in \cref{fig:pmcal_cmd_pops_pm_a}, distinct 47 Tuc and SMC populations can also be seen in \cref{fig:pmcal_cmd_pops_pm_b}.
Most of the 47 Tuc stars have $\mu < 2.5~\mathrm{mas}~\mathrm{yr}^{-1}$, while the total proper motions of the SMC stars are clustered near $\mu \sim 5~\mathrm{mas}~\mathrm{yr}^{-1}$.
Note that the density of stars at $\mu = 0$ is approximately zero due to the factor of $\mu$ that appears through the Jacobian in transforming the proper motion distribution function from Cartesian coordinates to polar coordinates, as explained in \cref{sec:deepacs_pmcal_pm_dist}.

In terms of the total proper motion given by \cref{eq:total_pm}, the 47 Tuc proper motion cut selects objects with $\mu < 2.5~\textrm{mas}~\textrm{yr}^{-1}$, which corresponds to all objects that lie below the solid black line in \cref{fig:pmcal_cmd_pops_pm_b}.
The SMC proper motion cut is more complicated in terms of $\mu$.
The distance between the mean proper motion of the SMC and the mean proper motion of 47 Tuc in proper motion space is $\bar{\mu}_s = \sqrt{\bar{\mu}_{\alpha,s}^2 + \bar{\mu}_{\delta,s}^2}$, which has the value $\bar{\mu}_s = 5.02~\mathrm{mas}~\mathrm{yr}^{-1}$. 
So the smallest value of $\mu$ along the boundary of the SMC proper motion cut is $\bar{\mu}_s - 0.75~\mathrm{mas}~\mathrm{yr}^{-1} = 4.27~\mathrm{mas}~\mathrm{yr}^{-1}$, and the largest value of $\mu$ along the boundary of the SMC proper motion cut is $\bar{\mu}_s + 0.75~\mathrm{mas}~\mathrm{yr}^{-1} = 5.77~\mathrm{mas}~\mathrm{yr}^{-1}$.
Thus, all objects selected by the SMC proper motion cut have total proper motion in the range $4.27~\mathrm{mas}~\mathrm{yr}^{-1} < \mu < 5.77~\mathrm{mas}~\mathrm{yr}^{-1}$; however, not all objects with $\mu$ in this range are actually selected by the SMC proper motion cut.

The increasing uncertainty of the proper motion measurements with increasing magnitude causes fainter stars in both 47 Tuc and the SMC to appear more dispersed, which manifests in \cref{fig:pmcal_cmd_pops_pm_b} as the increasingly large spread of $\mu$ values along the y-axis as the magnitude increases.
This leads to the proper motion distributions of the two populations overlapping more as the magnitude increases.
The number of SMC contaminants in the 47 Tuc white dwarf data space is thus expected to increase with magnitude, especially for $\text{F606W} \gtrsim 27$ where the proper motions start to become noticeably more dispersed and the CMD sequences of the SMC stars and 47 Tuc white dwarfs begin to intersect.

The value of $\Ncontam$ that we calculate in each F606W magnitude bin is given in \cref{tab:pmcal_completeness}. 
For reference, the total number of objects, $N_\mathrm{WD}$, found in that bin for our proper-motion-cleaned white dwarf data space (i.e. after both the 47 Tuc proper motion cut and the 47 Tuc white dwarf CMD cut) is also given in \cref{tab:pmcal_completeness}.
The estimate of the true number of white dwarfs in that bin is $N_\mathrm{WD} - \Ncontam$.
Since $\Ncontam$ is really an upper limit on the number of SMC contaminants in the 47 Tuc white dwarf data space (which should also be close to the actual number of contaminants), the quantity $N_\mathrm{WD} - \Ncontam$ is correspondingly really a lower limit on the true number of white dwarfs.
Note that the interpretation of $\Ncontam$ as an upper limit makes it sensible to report non-integer values for $\Ncontam$.

\begin{table}
    \centering
    \caption{Results of calibrating proper motion cleaning procedure by F606W magnitude bin.
    The number of objects in the proper-motion-cleaned 47 Tuc white dwarf data space ($N_\mathrm{WD}$) is the total number of objects in a given magnitude bin that survive both the 47 Tuc proper motion cut and the 47 Tuc white dwarf CMD cut.
    The number of contaminants ($\Ncontam$) is the estimated number of SMC stars that survive the same cuts used to calculate $N_\mathrm{WD}$.
    The completeness reduction factor ($f_\mathrm{CR}$) is the fraction of CMD-selected 47 Tuc main-sequence stars that survive the 47 Tuc proper motion cut.
    The error for $f_\mathrm{CR}$ is reported in the final column, following its value.
    All of these quantities were calculated using the \sharpparam-cleaned data.}
    \setlength\tabcolsep{5pt} %% default is 6pt (which is slightly too big here)
    \begin{tabularx}{\columnwidth}{>{\centering\arraybackslash}X >{\centering\arraybackslash}X >{\centering\arraybackslash}X >{\centering\arraybackslash}X >{\centering\arraybackslash}X}
        \toprule
        F606W & $N_\mathrm{WD}$ & $\Ncontam$ & $f_\mathrm{CR}$ & $\mathrm{Error}(f_\mathrm{CR})$\\
        \midrule
        $22.0 - 22.5$ &   $1$ & $0.0$ & $0.9413$ & $0.0062$\\
        $22.5 - 23.0$ &   $1$ & $0.0$ & $0.9373$ & $0.0058$\\
        $23.0 - 23.5$ &   $5$ & $0.0$ & $0.9349$ & $0.0055$\\
        $23.5 - 24.0$ &  $10$ & $0.0$ & $0.9359$ & $0.0050$\\
        $24.0 - 24.5$ &  $28$ & $0.0$ & $0.9431$ & $0.0051$\\
        $24.5 - 25.0$ &  $27$ & $0.0$ & $0.9328$ & $0.0060$\\
        $25.0 - 25.5$ &  $44$ & $0.0$ & $0.9293$ & $0.0069$\\
        $25.5 - 26.0$ &  $48$ & $0.0$ & $0.9114$ & $0.0088$\\
        $26.0 - 26.5$ &  $75$ & $0.0$ & $0.9093$ & $0.0106$\\
        $26.5 - 27.0$ &  $78$ & $0.0$ & $0.9087$ & $0.0131$\\
        $27.0 - 27.5$ & $117$ & $0.1$ & $0.8849$ & $0.0167$\\
        $27.5 - 28.0$ & $150$ & $0.0$ & $0.8161$ & $0.0240$\\
        $28.0 - 28.5$ & $320$ & $0.6$ & $0.7419$ & $0.0351$\\
        $28.5 - 29.0$ & $424$ & $4.5$ & $0.7016$ & $0.0411$\\
        \bottomrule
    \end{tabularx}
    \label{tab:pmcal_completeness}
\end{table}

From the values tabulated in \cref{tab:pmcal_completeness}, it can be seen that $\Ncontam$ is negligible compared to $N_\mathrm{WD}$ for our choice of 47 Tuc proper motion cut.
Most magnitude bins contain no contaminants, and even for the faintest magnitudes, most of the bins have $\Ncontam < 1$.
The largest number of contaminants is found in the faintest magnitude bin, $28.5 < \text{F606W} < 29.0$, and $\Ncontam$ is still negligible compared to $N_\mathrm{WD}$ for this bin.
As our cleaning procedure successfully removes all but a negligible number of SMC stars from the data, we do not need to apply a correction in our unbinned likelihood analysis to account for SMC contaminants in the white dwarf data space.
However, the proper motion cleaning also removes some objects that are actually 47 Tuc white dwarfs.
This reduces the completeness of our white dwarf sample below what is found from the artificial stars tests.
This completeness reduction effect is analysed below in \cref{sec:deepacs_pmcal_comp_red}, and the result of that analysis is also included in \cref{tab:pmcal_completeness}.

\subsubsection{Completeness Reduction}  \label{sec:deepacs_pmcal_comp_red}

The reduction in completeness from the proper motion cut to select likely 47 Tuc members is quantified using 47 Tuc main-sequence stars. 
These main-sequence stars are identified in the CMD using the same bounding region as was used for the \sharpparam\ calibration, which is fully shown in \cref{fig:sharpcal_pmcal_cmd_a} as the right-most bounding region (in green).
The proper motion cut used in our ultimate cleaning procedure for the white dwarf data, which selects objects with a total proper motion $< 2.5~\textrm{mas}~\textrm{yr}^{-1}$ relative to the mean proper motion of 47 Tuc, was then applied to the CMD-selected 47 Tuc main-sequence sample.
The fraction of main-sequence stars remaining after the proper motion cut quantifies the completeness reduction due to the proper motion cleaning procedure.
This survival fraction corresponds to the magnitude-dependent fraction of green points that lie below the solid black line in \cref{fig:pmcal_cmd_pops_pm_b} and tends to decreases with magnitude as the proper motion uncertainties increase.
The completeness reduction was determined as a function of F606W magnitude by sorting the main-sequence sample into bins of 0.5 magnitude width over the magnitude range 22 to 29 and calculating the fraction remaining in each bin after the proper motion cut.
These are the same magnitude bins used to calculate $\Ncontam$.
The results are tabulated in \cref{tab:pmcal_completeness}.

Let $\Ncm$ be the number of CMD-selected 47 Tuc main-sequence stars without any proper motion cut applied, and let $\Nptcm$ be the number of CMD-selected 47 Tuc main-sequence stars that survive the 47 Tuc proper motion cut.
We define a completeness reduction factor $f_\mathrm{CR}$, which is given by the survival fraction of the CMD-selected main-sequence stars
\begin{equation}
    f_\mathrm{CR} = \frac{\Nptcm}{\Ncm}.
\end{equation}
The standard error in $f_\mathrm{CR}$ is taken to be binomially-distributed and thus given by
\begin{equation}
    \mathrm{Error}(f_\mathrm{CR}) = \sqrt{\frac{f_\mathrm{CR} \, (1 - f_\mathrm{CR})}{\Ncm}}.
\end{equation}
The errors calculated in this way are reported in the final column of \cref{tab:pmcal_completeness}, following the corresponding values of $f_\mathrm{CR}$.

\begin{figure}
    \centering
    \includegraphics[width=\columnwidth]{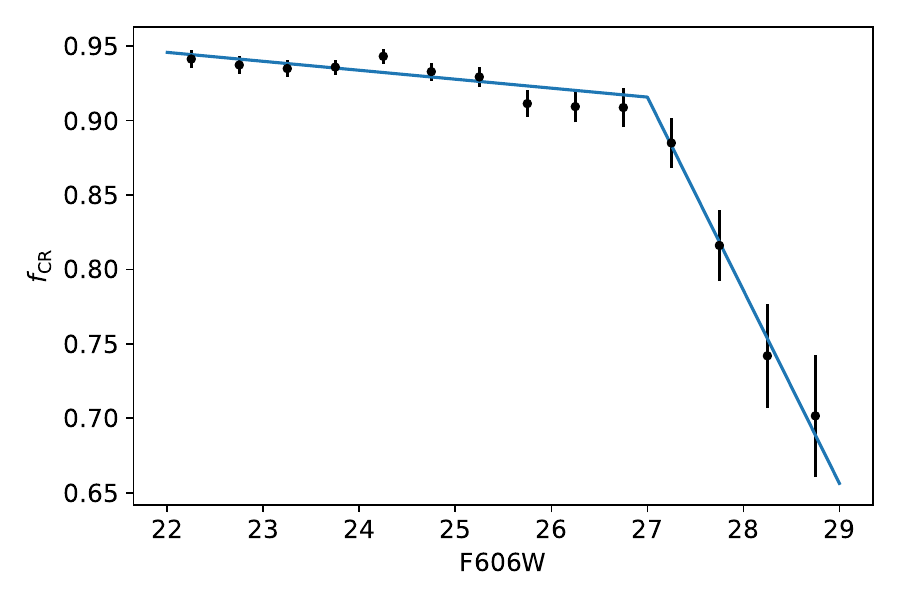}
    \caption{Completeness reduction factor $f_\mathrm{CR}$ due to proper motion cleaning as a function of F606W magnitude. 
    The analytic function (blue curve) is shown for the best-fitting parameters determined by fitting the binned values of $f_\mathrm{CR}$ (black points) calculated in the calibration of the cleaning procedure.}
    \label{fig:deepacs_fcr_fit}
\end{figure}

To avoid numerical artefacts due to binning when using $f_\mathrm{CR}$ in the unbinned likelihood analysis, we modelled $f_\mathrm{CR}$ as a piece-wise linear function of F606W with parameter values determined by fitting this function to the reference values given in \cref{tab:pmcal_completeness}.
The analytic function used for $f_\mathrm{CR}$ consists of two linear segments, with the switch occurring at the F606W value $\mathrm{F606W}_0$ and corresponding $f_\mathrm{CR}$ value $f_{\mathrm{CR},0}$.
We let $\mathrm{F606W}_0$ and $f_{\mathrm{CR},0}$ be free parameters of the fit, along with the slope $a_1$ of the segment where $\mathrm{F606W} < \mathrm{F606W}_0$ and the slope $a_2$ of the other segment where $\mathrm{F606W} \geq \mathrm{F606W}_0$.
The best-fitting parameter values are given in \cref{tab:deepacs_fcr_fit_params}, and the corresponding best-fitting function is plotted in \cref{fig:deepacs_fcr_fit}, along with the values it was fitted to.
The reduced chi-squared $\chi_\nu$ value (with $\nu = 10$ degrees of freedom) is $\chi_\nu = 0.976$, indicating a good fit.

\begin{table}
    \centering
    \caption{Best-fitting parameter values for piece-wise linear model of $f_\mathrm{CR}$ as a function of F606W.}
    \begin{tabularx}{\columnwidth}{X c X D{+}{\:\pm\:}{6} X}
        \toprule
        ~ & Parameter & ~ & \heading{Value} & ~\\
        \midrule 
        ~ & $\mathrm{F606W}_0$ & ~ & 27.00 + 0.17 & ~\\
        ~ & $f_{\mathrm{CR},0}$ & ~ & 0.9158 + 0.0059 & ~\\
        ~ & $a_1$ & ~ & -0.0060 + 0.0017 & ~\\
        ~ & $a_2$ & ~ & -0.1297 + 0.0247 & ~\\
        \bottomrule
    \end{tabularx}
    \label{tab:deepacs_fcr_fit_params}
\end{table}

The completeness reduction due to proper motion cleaning is accounted for in our unbinned likelihood analysis after applying the photometric error distribution function, \cref{eq:deepacs_errdist}, to the cooling model.
Note that the error distribution is taken to depend on the input magnitudes, whereas the completeness reduction factor should be treated as a function of the output magnitudes after accounting for photometric errors (as $f_\mathrm{CR}$ was calculated directly from observations).
After applying the error distribution to the cooling model, the resultant distribution function (which is a function of the output magnitudes) is then multiplied by $f_\mathrm{CR}$ to account for the completeness reduction in the unbinned likelihood analysis.

\section{Models} \label{sec:deepacs_models}

We created white dwarf cooling models using the stellar evolution software MESA \citep{mesa1,mesa2,mesa3,mesa4,mesa5}\footnote{There is also a more recent sixth instrument paper \citep{mesa6}; however, it is only relevant for MESA versions released later than the most recent version used in this work.}. 
Using MESA, we ran a suite of white dwarf cooling simulations for different parameter values, varying the white dwarf mass and the thickness of the H envelope.
This suite of simulations was generated from an initial model of a young, hot white dwarf that was created by simulating the pre-white dwarf evolution of a progenitor star. 
The simulation that produced this initial model is described in \cref{sec:deepacs_models_initial_model}. 
The white dwarf cooling models are described in \cref{sec:deepacs_models_wd_cooling_models}.

\subsection{Creation of Initial Model} \label{sec:deepacs_models_initial_model}

The initial model for the white dwarf cooling simulations was created by simulating the evolution of a 0.9 $M_\odot$ progenitor star with parameters appropriate for 47 Tuc from the pre-main sequence until the birth of the white dwarf.
This simulation was done using MESA version 10398 (mesa-r10398) and was created from the MESA \texttt{test\_suite} example \texttt{1M\_pre\_ms\_to\_wd}.
We modified the parameters of the \texttt{1M\_pre\_ms\_to\_wd} inlist by changing the initial mass, the initial composition parameters, and the wind parameters. We set the following initial parameters
\begin{verbatim}
    initial_mass = 0.9d0
    initial_z = 4.0d-3
    initial_y = 0.256d0
\end{verbatim}
and set \texttt{Zbase} (for use with Type 2 opacities) to be the same as \texttt{initial\_z}.

We used a Reimers mass loss scheme \citep{1975MSRSL...8..369R} on the red giant branch (RGB) and a Blocker mass loss scheme \citep{1995A&A...297..727B} on the asymptotic giant branch (AGB).
Previous work has shown that stars in 47 Tuc lose most of their mass on the AGB \citep{2015ApJ...810..127H}, rather than the RGB, so we set the scaling parameter values
\begin{verbatim}
    Reimers_scaling_factor = 0.1d0
    Blocker_scaling_factor = 0.7d0
\end{verbatim}
for our prescriptions of mass loss via stellar winds. 
As we are not primarily concerned with the details of stellar evolution before the white dwarf stage, the particular choice of scaling factors is not a concern for this work; the values simply need to be reasonable for 47 Tuc and produce a white dwarf with a thick H envelope from which our set of white dwarf models can be created.

A custom stopping condition was used to ensure that the simulation ended shortly after the star became a white dwarf. The simulation ended when the two conditions \texttt{log\_Teff} $> 4.5$ and \texttt{log\_L} $< 2$ were both met, with the code for this implemented through the \texttt{run\_star\_extras} module.
This stopped the simulation before the luminosity of the white dwarf had dropped into the luminosity range of interest for studying white dwarf cooling in 47 Tuc, and thus the output of this simulation can be used as a starting point to generate white dwarf cooling models that span the entire luminosity range of interest.
In the end, the simulation of pre-main sequence to white dwarf evolution produced a model of a newly born white dwarf with a mass of $0.5388~M_\odot$ and a thick H envelope.

\begin{figure}
    \centering
    \begin{subfigure}{\columnwidth}
        \centering
        \includegraphics[width=\columnwidth]{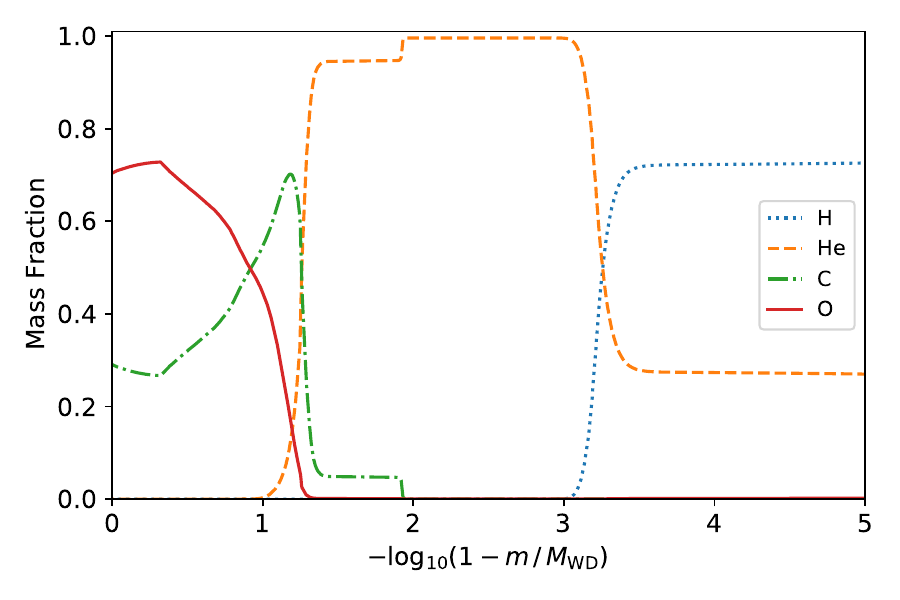}
        \caption{Profile at the start of white dwarf cooling.}
        \label{fig:composition_profile_early}
    \end{subfigure}\\[1ex]
    \begin{subfigure}{\columnwidth}
        \centering
        \includegraphics[width=\columnwidth]{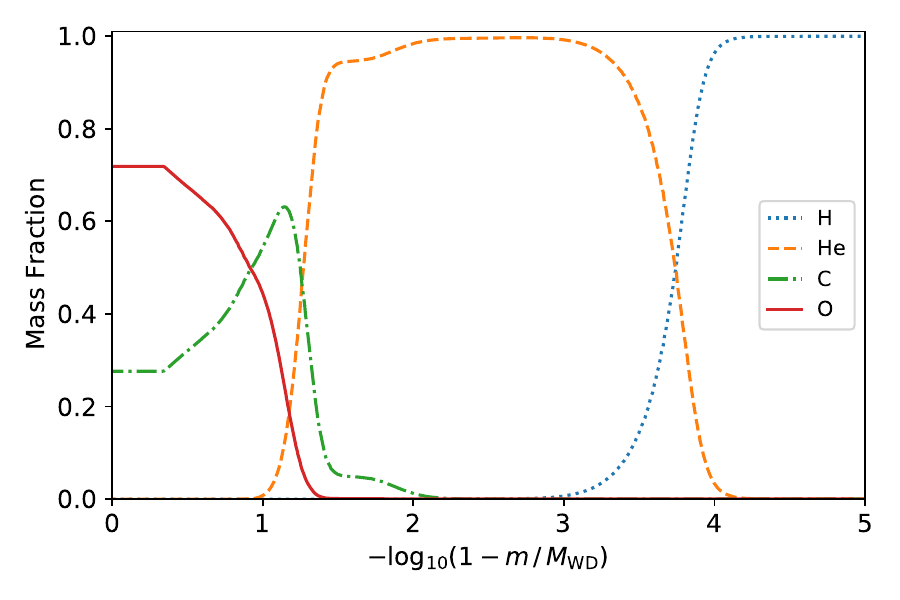}
        \caption{Profile after $1~\mathrm{Gyr}$ of white dwarf cooling.}
        \label{fig:composition_profile_late}
    \end{subfigure}\\[1ex]
    \caption{Composition profiles showing mass fractions of key elements for a $0.5388~M_\odot$ white dwarf model with an envelope thickness parameter of $\lqh = -3.55$.
    The top panel shows the composition profile near the start of the white dwarf cooling simulations (when the effective temperature is approximately $85,000~\mathrm{K}$), while the bottom panel shows a typical composition profile after $1~\mathrm{Gyr}$ of cooling (when the effective temperature is approximately $8,000~\mathrm{K}$).}
    \label{fig:composition_profile}
\end{figure}

The composition profile produced by this initial simulation is shown in \cref{fig:composition_profile_early}, where the mass fractions are plotted for the key elements comprising the white dwarf.
The profile corresponds to a $0.5338~M_\odot$ white dwarf model at the beginning of the white dwarf simulations described in \cref{sec:deepacs_models_wd_cooling_models} when the effective temperature is approximately $85{,}000~\mathrm{K}$.
The peak in oxygen abundance occurring just outside the centre of the white dwarf is due to non-nuclear neutrino processes that occur at the very early stage of white dwarf cooling, most significantly the plasma neutrino process (and to a lesser extent neutrino bremsstrahlung and the photoneutrino process).
The temperature likewise peaks just outside the centre of the white dwarf at this very early cooling stage.
At later stages of white dwarf cooling, this feature is flattened out so that the maximum oxygen abundance occurs in the centre of the white dwarf, with the oxygen abundance remaining constant at this maximum value for the innermost part of the profile within the C/O core.
Diffusion in our white dwarf cooling simulations, which will be described in \cref{sec:deepacs_models_wd_cooling_models} below, also quickly leads to the composition profile becoming more stratified, with the outer envelope becoming nearly pure H.

For comparison, a typical composition profile at a later stage of white dwarf evolution is shown in \cref{fig:composition_profile_late}.
The compositions profiles in \cref{fig:composition_profile_early,fig:composition_profile_late} are shown for the same white dwarf model with a white dwarf mass of $\mwd = 0.5388~M_\odot$ and envelope thickness parameter (as defined in \cref{sec:deepacs_models_wd_cooling_models}) of $\lqh = -3.55$, just at different cooling times.
\cref{fig:composition_profile_late} is shown in particular for a cooling age of $1~\mathrm{Gyr}$, at which time the effective temperature is approximately $8{,}000~\mathrm{K}$; however, this plot is representative of the composition for most of the period of white dwarf cooling in our suite of cooling simulations for various model parameters.
\cref{fig:composition_profile_late} displays the flattened inner core of the oxygen abundance profile and further stratification of the element layers noted above.
For the plotted model, the mass fraction of oxygen at the centre of the white dwarf is initially $X_\mathrm{centre}(\mathrm{O}) = 0.70$ and reaches an approximately constant value of $X_\mathrm{centre}(\mathrm{O}) = 0.72$ by a cooling age of $10^7~\mathrm{yrs}$.
The average mass fractions of oxygen and carbon over the whole white dwarf, respectively denoted $X_\mathrm{av}(\mathrm{O})$ and $X_\mathrm{av}(\mathrm{C})$, are constant over time throughout the evolution of the white dwarf with values of $X_\mathrm{av}(\mathrm{O}) = 0.63$ and $X_\mathrm{av}(\mathrm{C}) = 0.31$.
The average mass fraction of He in the white dwarf, which parametrises the relative mass of the He layer once the element layers have become more stratified, is approximately constant at a value of $X_\mathrm{av}(\mathrm{He}) = 5.4 \times 10^{-2}$ (with the He content produced by residual H burning in the envelope being largely negligible compared to this value).
These composition parameter values are typical for our suite of white dwarf cooling models because the models are generated from the same initial model that is well-motivated by the simulation of the pre-white dwarf stages of evolution.

\subsection{White Dwarf Cooling Models} \label{sec:deepacs_models_wd_cooling_models}

Starting from the model created by the simulation described in \cref{sec:deepacs_models_initial_model}, we generated additional initial models with different masses and envelope thicknesses from which to begin the white dwarf cooling simulations.
We then simulated the evolution of each of these initial white dwarf models over time to create a suite of white dwarf cooling models.
Both the procedures to modify the initial model and the main white dwarf cooling simulations were performed using MESA version 15140 (mesa-r15140).

From the initial model of a $0.5338~M_\odot$ white dwarf, we created less massive white dwarf models with masses of 0.5092, 0.5166, 0.5240, and 0.5314~$M_\odot$ using the \texttt{relax\_mass\_scale} control provided by MESA and running a brief simulation in which the model was allowed to evolve for a few short time steps to adjust to the change. 
Note that a mass of $0.5240~M_\odot$ was chosen as one of the target mass values because this was the mass of the model used in \citet{Obertas2018}, which was produced by a stellar evolution simulation of a $0.974~M_\odot$ progenitor for 47 Tuc and was found by \citet{Obertas2018} to reasonably replicate the cooling curve of the old white dwarfs in the same data as used in this work.
The procedure to reduce the mass of the initial model simply re-scaled the profile of the $0.5338~M_\odot$ input model to the target mass, so like the $0.5338~M_\odot$ model, the less massive models produced in this way also have thick H envelopes.

We found that the \texttt{relax\_mass\_scale} procedure was only able to successfully re-scale models to a target mass that was lower than the mass of the initial model.
In order to extend the mass grid to masses above 0.5338~$M_\odot$, we thus created another, heavier initial model from which the simulations for more massive white dwarfs could be generated.
We created a 0.5644~$M_\odot$ initial white dwarf model using a simulation based on the \texttt{test\_suite} example \texttt{make\_co\_wd} with the relevant parameters modified to make the simulation appropriate for 47 Tuc.
From the new 0.5644~$M_\odot$ initial white dwarf model, we used the \texttt{relax\_mass\_scale} procedure described above to create models with white dwarf masses of 0.5462 and 0.5536~$M_\odot$.
We also checked to confirm that relaxing the heavier initial white dwarf model down to lower masses produced the same cooling models as starting from the $0.5388~M_\odot$ initial model.
In total, we created white dwarf simulations for white dwarf masses spanning the range $0.5092 - 0.5535~M_\odot$ (inclusive) in increments of $0.0074~M_\odot$.

For each of these masses, models with thinner envelopes were created by using the \texttt{relax\_mass} control. 
This removes mass from the white dwarf via a wind, which takes the mass from the H envelope. 
For this procedure the new mass is set to be just slightly smaller than the mass of the input model, with the difference being the amount of mass to remove from the H envelope. 
As even the thickest envelopes only have a mass on the order of $10^{-4}$ times the total mass $M_\mathrm{WD}$ of the white dwarf, the change to $M_\mathrm{WD}$ due to reducing the mass of the envelope in this way is negligible in terms of the effect that varying $M_\mathrm{WD}$ can have on the cooling curves.
For white dwarfs with thick envelopes, the envelope thickness also decreases over time at early times in the white dwarf's evolution due to residual H burning near the boundary of the He layer and H envelope. 
This residual H burning does not change the total mass of the white dwarf, but it makes the envelope thickness in general a function of time.

To define a parameter quantifying the thickness of the H envelope, we select a reference cooling time of $10~\mathrm{Myr}$ into the white dwarf cooling simulations at which to define the envelope thickness parameter, and for this parameter we use the relative mass of the H envelope $q_H = M_H / M_\mathrm{WD}$, where $M_H$ is the mass of the H envelope.
It should be noted that when performing simulations for different initial envelope thicknesses, we found that there is a physical upper limit on how large $q_H$ can be. For a sufficiently large initial envelope thickness (i.e. at a cooling time of $0$~yrs), any additional H that initially makes the envelope thicker is quickly burnt away through residual H burning on time scales earlier than what is relevant for our work, resulting in the same envelope thickness by the reference time of $10~\mathrm{Myr}$ at which $q_H$ is defined.

\begin{figure*}
    \centering
    \begin{subfigure}[t]{0.49\textwidth}
        \centering
        \includegraphics[width=\textwidth]{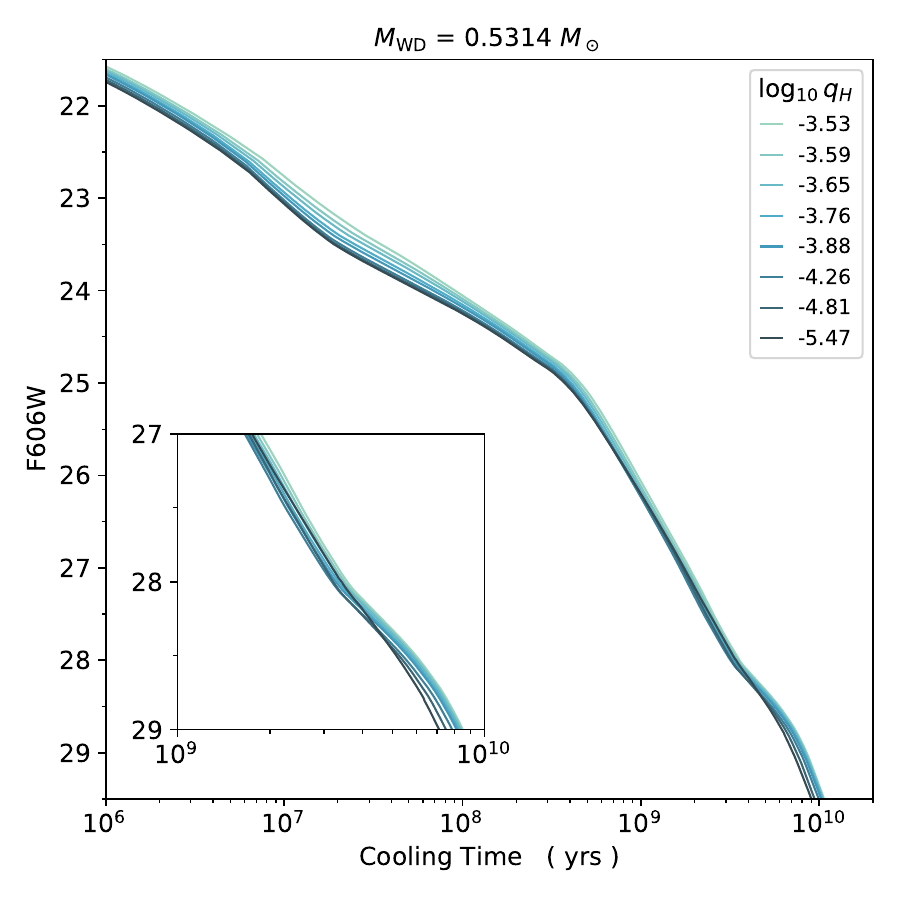}
        \caption{Varying $\lqh$ for fixed $\mwd$.}
        \label{fig:cooling_curves_qhseries}
    \end{subfigure}
    \hfill
    \begin{subfigure}[t]{0.49\textwidth}
        \centering
        \includegraphics[width=\textwidth]{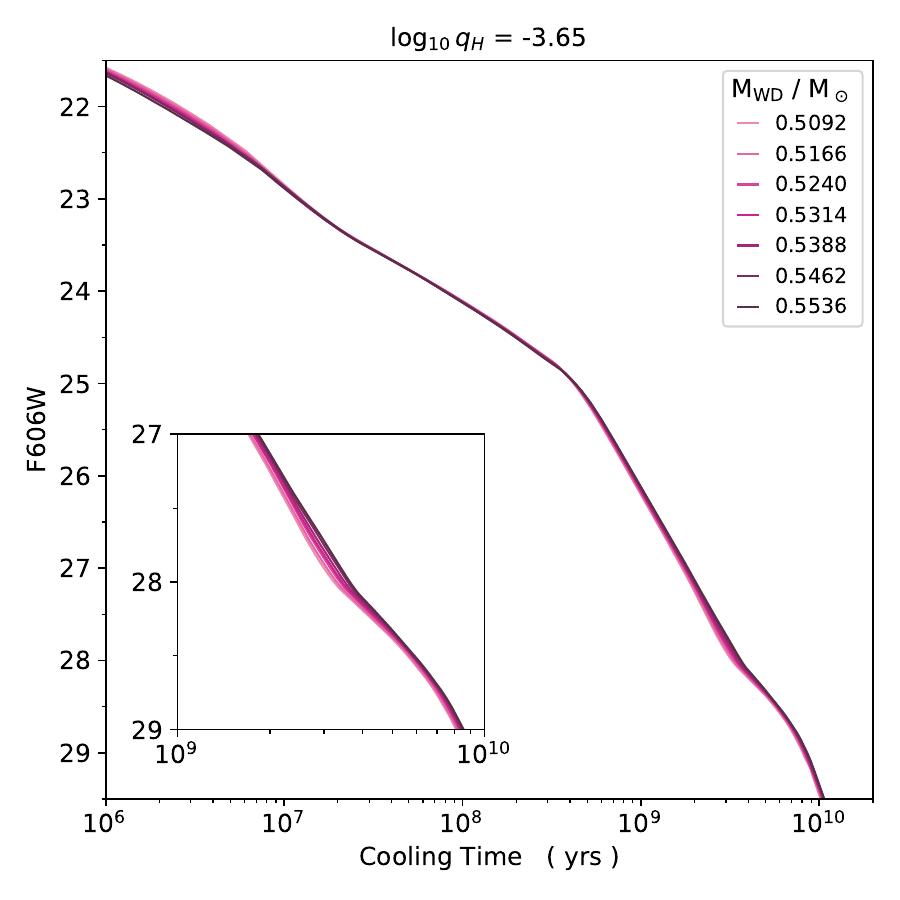}
        \caption{Varying $\mwd$ for fixed $\lqh$.}
        \label{fig:cooling_curves_mwdseries}
    \end{subfigure}
    \caption{Theoretical cooling curves for different model parameters varied in the MESA simulations. The left panel shows cooling curves for different envelope thicknesses when the white dwarf mass is fixed at $\mwd = 0.5314~M_\odot$, while the right panel shows cooling curves for different white dwarf masses when the envelope thickness is fixed at $\lqh = -3.65$. The inset plots focus on the bump associated with convective coupling of the white dwarf envelope to the core.%
    }
    \label{fig:cooling_curves}
\end{figure*}

The cooling models produced by these simulations describe how luminosity changes over time, and the cooling curves showing this relation for the different parameter combinations that were simulated are plotted in \cref{fig:cooling_curves}. 
For these cooling curves, the luminosity has been converted to the apparent magnitude in the F606W filter using the same procedure as used in the unbinned likelihood analysis and described in \cref{sec:deepacs_unbinned_likelihood}.
\cref{fig:cooling_curves_qhseries} shows the effect of varying the envelope thickness for a fixed white dwarf mass of $\mwd = 0.5314~M_\odot$, while \cref{fig:cooling_curves_mwdseries} shows the effect of varying the white dwarf mass for a fixed envelope thickness of $\lqh = -3.65$.
Though the difference between the curves for different $\mwd$ values in \cref{fig:cooling_curves_mwdseries} is less visually distinct than for the curves of different $\lqh$ values in \cref{fig:cooling_curves_qhseries}, our unbinned likelihood analysis of the data studied in this work is still sensitive enough to distinguish between them, which will be seen in the results presented in \cref{sec:deepacs_results}.
Note that the curves shown in \cref{fig:cooling_curves_qhseries} correspond to the original models produced by MESA without interpolating.
In order to compare models of different $\mwd$ values in the unbinned likelihood analysis (and in \cref{fig:cooling_curves_mwdseries}), the cooling curves were interpolated onto a grid of $\lqh$ values.

The final bump in the cooling curve that occurs after a few billion years of cooling time in \cref{fig:cooling_curves} (just below a magnitude value of 28 in the plots) corresponds to the onset of convective coupling of the envelope to the core, which also approximately coincides with the onset of core crystallisation for white dwarfs in 47 Tuc \citep{Obertas2018}.
This feature is sensitive to the envelope thickness and occurs at a luminosity where there are a large number of white dwarfs in the deep ACS data set.
In \cref{fig:cooling_curves_qhseries}, it can be seen that the bump associated with convective coupling is reduced for thinner envelopes. In the case of the thinnest envelope shown, the bump is nearly completely flattened out, with the part of the curve before the bump being raised in addition to the luminosity of the bump being lowered.

\section{Birthrate} \label{sec:deepacs_birthrate}

The white dwarf birthrate for our sample is one of the parameters that is determined in the unbinned likelihood analysis described in \cref{sec:deepacs_unbinned_likelihood}. Before performing that analysis, however, we first determine a prior for the birthrate using observations of red giant stars from \gaia\ EDR3 data in the HST footprint.
Since the members of a star cluster population are approximately the same age and the stars evolve quickly through the evolutionary stages between the end of the main sequence stage and start of the white dwarf stage, the rate of stars leaving the main sequence should be approximately the same as the white dwarf birthrate, and this rate can be measured using stars on the RGB.

The HST data cannot be used for this birthrate calculation because the red giant stars are so bright that they saturate the deep observations, leading to high incompleteness.
Instead, we must use another dataset and ensure that we select stars over the same field of view as was used for the HST white dwarf observations. 
\gaia\ EDR3 observations are essentially complete for $17 < G < 12$, which spans nearly the entire RGB of 47 Tuc, and have the most precise astrometry ever measured\footnote{Though the more recent \gaia\ DR3 is now available, the astrometric measurements are the same as \gaia\ EDR3.}, making \gaia\ EDR3 a good dataset for this purpose.

\begin{figure*}
    \centering
    \begin{subfigure}{0.49\textwidth}
    \centering
    \includegraphics{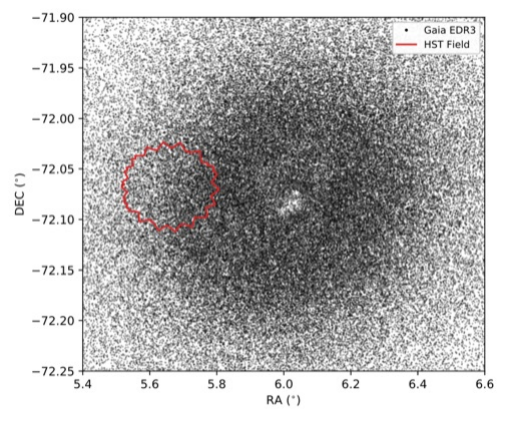}
    \caption{Zoomed-out view before selection.}
    \label{fig:deep_acs_birthrate_field_a}
    \end{subfigure}
    \begin{subfigure}{0.49\textwidth}
    \centering
    \includegraphics{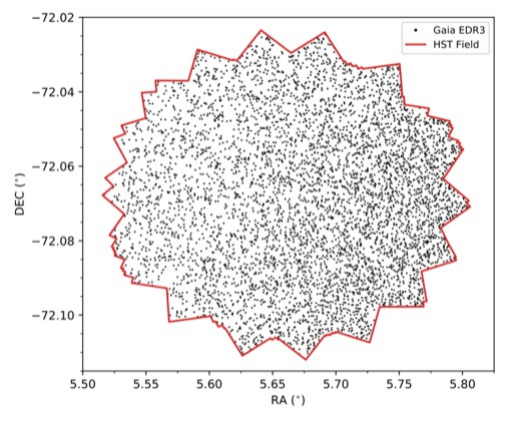}
    \caption{Zoomed-in view after selection.}
    \label{fig:deep_acs_birthrate_field_b}
    \end{subfigure}
    \caption{Field boundaries (red curves) for the HST ACS/WFC deep observations overlaid on \gaia\ EDR3 observations (black points) of 47 Tuc.
    The \gaia\ EDR3 sources that are located within these boundaries are used for the birthrate calculations.}
    \label{fig:deep_acs_birthrate_field}
\end{figure*}

We first retrieved all \gaia\ EDR3 sources within a radius\footnote{A value of $5^\circ$ was used for this radius, but the exact value is not important as long as it is large enough to include the entire HST field of view.} far enough from the centre of 47 Tuc that the selection included the entire HST field of view. 
The \gaia\ data is publicly available and was retrieved through Vizier.
We then selected the \gaia\ EDR3 sources within the field boundaries of the HST ACS/WFC observations of our white dwarf data by performing a cut in position space to get a sample of \gaia\ EDR3 sources in the HST footprint. 
To get the field boundary for the full HST observation, we merged the boundaries of all of the orbits.
The observational plan grouped the orbits into 24 visits, and the boundaries for the 24 visits (for which the observing regions of the constituent orbits%
\footnote{Each visit consists of a group of five orbits (except for Visit 24, which consists of six orbits). See \citet{2012AJ....143...11K} for details.} %
have been merged) are publicly available through MAST.
We merged the boundaries provided by MAST for these visits to get the boundary for the final stacked image that combined the observations from all of the orbits. 
The final merged boundary is shown as a red curve in \cref{fig:deep_acs_birthrate_field}, where it is plotted over the \gaia\ EDR3 data (shown as the black points).
\Cref{fig:deep_acs_birthrate_field_a} shows the \gaia\ EDR3 data before selecting sources within the HST field boundary; this plot is approximately centred on the centre of 47 Tuc and shows the location of the HST observations relative to the cluster centre.
\Cref{fig:deep_acs_birthrate_field_b} shows only the \gaia\ EDR3 sample after selecting sources within the HST field boundary, with the plot approximately centred on the centre of the HST field boundary.

The dependence of the number density and completeness on the radial distance from the cluster centre is clearly shown in \cref{fig:deep_acs_birthrate_field_a}.
The number density in general increases as the distance to the cluster centre decreases; however, overcrowding near the very centre of the cluster notably reduces the completeness in this region, leading to an absence of stars observed at the centre where the cluster is most densely populated.
The dependence of the completeness and thus photometric error distribution on the radial coordinate from the cluster centre would be a concern if the region observed to gather the white dwarf data were closer to the centre; however, the field of view for the HST observations considered in this work is far enough away from the cluster centre that the photometric error distribution does not depend appreciably on radius.
The dependence of the number density on radius can also be seen in \cref{fig:deep_acs_birthrate_field_b}, where the number density of the data points increases with increasing values of right ascension, which correspond to positions closer to the cluster centre within the HST field region shown in that plot.
Since the photometric error distribution does not depend on radius, the dependence of the number density on radius does not need to be accounted for in the unbinned likelihood analysis.

\begin{figure*}
    \centering
    \begin{subfigure}{0.49\textwidth}
    \centering
    \includegraphics{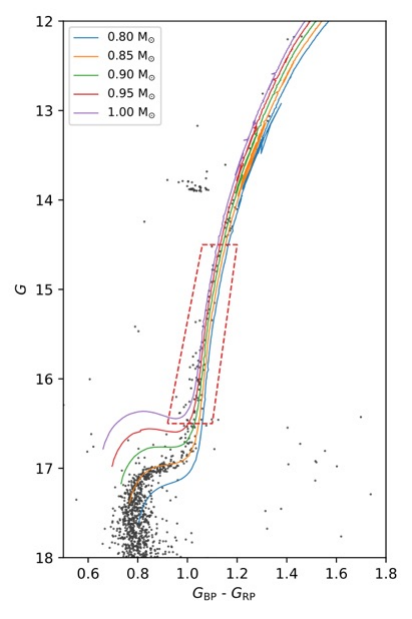}
    \caption{Selection excluding red-giant bump.}
    \label{fig:deep_acs_birthrate_rgb_selection_short}
    \end{subfigure}
    \begin{subfigure}{0.49\textwidth}
    \centering
    \includegraphics{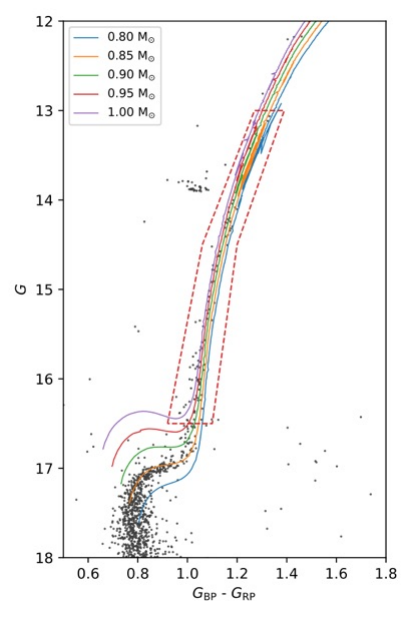}
    \caption{Selection including red-giant bump.}
    \label{fig:deep_acs_birthrate_rgb_selection_long}
    \end{subfigure}
    \caption{CMD selections of RGB stars for birthrate calculations. 
    The black points correspond to \gaia\ EDR3 data in the HST footprint, while the dashed red curves indicate the boundaries to select the RGB stars.
    From right to left, the solid curves correspond to stellar evolution models with initial masses of 0.80, 0.85, 0.90, 0.95, and 1.00~$M_\odot$.}
    \label{fig:deep_acs_birthrate_cmd}
\end{figure*}

From the sample of \gaia\ EDR3 sources within the HST field boundaries, we then selected RGB stars using a cut in the CMD. 
The main boundary regions for selecting RGB stars is shown in \cref{fig:deep_acs_birthrate_cmd} by the dashed red curves.
\Cref{fig:deep_acs_birthrate_cmd} also shows stellar evolution models (solid curves) for various initial masses plotted over the \gaia\ EDR3 data (black points).
The models are shown in \cref{fig:deep_acs_birthrate_cmd} from the end of the main sequence stage (taken to be when the mass fraction of hydrogen at the centre of the star drops below $10^{-4}$) until the tip of the RGB.
The birthrate is calculated by dividing the number $N_\mathrm{RGB}$ of stars observed in the RGB boundary region by the time $t_\mathrm{RGB}$ it takes a star to traverse this region according to the models, so the (prior value for the) birthrate is given by $\dot{N}_0 = N_\mathrm{RGB} \, / \, t_\mathrm{RGB}$.
To assess the uncertainty in the birthrate, we calculated the birthrate using models of different initial masses and using different RGB boundary definitions consisting of sub-regions of the main boundary regions shown in \cref{fig:deep_acs_birthrate_cmd}.

The stellar evolution models used to calculate the birthrate (and shown in \cref{fig:deep_acs_birthrate_cmd}) were created by running MESA simulations analogous to the simulation described in \cref{sec:deepacs_models_initial_model} (which created the initial white dwarf model from which the white dwarf cooling simulations were generated) but with different initial masses.
Like the simulation of \cref{sec:deepacs_models_initial_model}, these simulations were based on the MESA \verb|test_suite| example \verb|1M_pre_ms_to_wd| with composition appropriate for 47 Tuc (i.e. with an \verb|initial_z| of $4 \times 10^{-4}$ and \verb|initial_y| of $0.256$) and were run using mesa-r10398.
The magnitudes (and colour) in the \gaia\ bandpass filters were calculated for these MESA evolution models using bolometric corrections.
We used the bolometric corrections%
\footnote{Accessed through \url{http://stev.oapd.inaf.it/cgi-bin/cmd} using version 3.4. This particular version is available at \url{http://stev.oapd.inaf.it/cgi-bin/cmd_3.4}.} %
of \citet{2019A&A...632A.105C} from PARSEC \citep{2012MNRAS.427..127B} calculated using synthetic spectra from a mix of the ATLAS9 ODFNEW \citep{2003IAUS..210P.A20C} and PHOENIX BT-Settl \citep{2012EAS....57....3A} spectral libraries, combined with the spectra from COMARCS \citep{2009A&A...503..913A,2016MNRAS.457.3611A} for cool giants and \citet{2015MNRAS.452.1068C} for very hot stars, with the transmission curves of the \gaia\ EDR3 bandpass filters \citep{2021yCat..36490003R,2021A&A...649A...3R} provided on the ESA/\gaia\ website\footnote{\url{https://www.cosmos.esa.int/web/gaia/edr3-passbands}}.
We used the same values for distance modulus and colour excess as are used in the unbinned likelihood analysis for the main white dwarf data, i.e. $\mu = 13.24$ and $E(B-V) = 0.04$, and we likewise used the same extinction curves, i.e. those of \citet{1989ApJ...345..245C} and \citet{1994ApJ...422..158O} with a total extinction%
\footnote{The total extinction is related to the colour excess through the relative visibility by the equation $A_V = R_V \, E(B-V)$, and a value of $R_V = 3.1$ is typical for the Milky Way.} %
of $A_V = 0.124$ and a relative visibility of $R_V = 3.1$.
The procedure for applying the bolometric corrections is essentially the same as what is described in \cref{sec:deepacs_unbinned_likelihood} for the white dwarf models; we simply used different filters and got the corresponding bolometric corrections from a different source for the RGB models used here compared to the white dwarf models used in the main analysis later.

The main models used for the birthrate calculations were those with initial masses of $0.85$, $0.90$, and $0.95$~$M_\odot$.
The models shown in \cref{fig:deep_acs_birthrate_cmd} indicate that stars with an initial mass of $\sim 0.85~M_\odot$ are just beginning to leave the main sequence, which is in keeping with expectations from other work \citep[e.g.][]{2003AJ....125..197H,2010AJ....139..329T,2016ApJ...816...70F,2016ApJ...826...88P}.
The $0.80~M_\odot$ model is only shown for reference to illustrate the mass-dependence of the main sequence turnoff; it was not used for any of the birthrate calculations because the models indicate that stars with mass this low in 47 Tuc have not yet left the main sequence.
The $1.00~M_\odot$ model was only used when calculating birthrates for truncated boundary regions that do not include its subgiant stage (i.e. boundary selection with a maximum value of $G \leq 16.4$).
The models with initial mass $\geq 0.85~M_\odot$ align well with the observed RGB sequence aside from the discrepancy between the location of the red-giant bump in the data and the corresponding feature in the models. 
The red-giant bump is the accumulation of stars located at $G \sim 14$ on the RGB of the data in \cref{fig:deep_acs_birthrate_cmd} and corresponds to a temporary decrease in luminosity (and thus a ``bump'' in the luminosity function).
The occurrence of this feature is related to the H-burning shell passing through the composition gradient left over by the convective envelope where it reached its maximum depth \citep{2015MNRAS.453..666C}.
The location of the red-giant bump is thus sensitive to the details of mixing processes beyond convective boundaries, which current stellar evolution models struggle to accurately predict \citep{2018ApJ...859..156K}.
The discrepancy between observations and model predictions of the luminosity of the red-giant bump in globular clusters has been identified repeatedly in the literature, with a variety of different potential explanations suggested \citep[e.g.][]{2006ApJ...641.1102B,2010ApJ...712..527D,2011A&A...527A..59C,2011PASP..123..879T,2015ApJ...814..142J,2018MNRAS.476..496F,2018ApJ...859..156K}, and it remains an open issue.
However, more accurately modelling the location of this feature is not needed for our purposes as long as the RGB boundary region is chosen appropriately.

The boundary region shown in \cref{fig:deep_acs_birthrate_rgb_selection_short} selects stars at the early stage of evolution along the RGB and excludes the red-giant bump feature of both the data and the models. 
This boundary region was chosen to maximise the number of stars selected while not needing to be concerned about the misalignment of the red-giant bump between the models and the data.
For each model used (as described above), we calculated $N_\mathrm{RGB} \, / \, t_\mathrm{RGB}$ for this region as well as sub-regions where the maximum $G$ value (at which the bottom horizontal boundary line in \cref{fig:deep_acs_birthrate_rgb_selection_short} was drawn) was reduced in increments of $0.1$ from $16.5$ to $16.0$.
This varies the boundary close to the subgiant branch, which stars evolve through more slowly than the RGB stage.
The spread in values from these calculations was taken to be the $3 \, \sigma$ range of the Gaussian prior for the birthrate (i.e. the range from $\dot{N}_0 - 3 \, \sigma_{\dot{N}}$ to $\dot{N}_0 + 3 \, \sigma_{\dot{N}}$), yielding a prior birthrate of $\dot{N}_0 = 2.21 \times 10^{-7}~\mathrm{yr}^{-1}$ with $\sigma_{\dot{N}} = 0.04 \times 10^{-7}~\mathrm{yr}^{-1}$.

As a check, we considered an extended RGB boundary region shown in \cref{fig:deep_acs_birthrate_rgb_selection_long} that includes the red-giant bump of both the data and the models%
\footnote{Using a boundary region that includes the red-giant bump in the data but excludes the corresponding feature in the models gives a larger birthrate estimate than regions that either include the red-giant bump in both the data and models or exclude it in both the data and models, but this is not a reasonable choice for the boundary region.} %
and performed a similar set of calculations in which the boundary regions of both \cref{fig:deep_acs_birthrate_rgb_selection_short} and \cref{fig:deep_acs_birthrate_rgb_selection_long} were truncated at the maximum $G$ end in increments of $0.25$ from $16.5$ to $16.0$, with $N_\mathrm{RGB} \, / \, t_\mathrm{RGB}$ calculated in each of these cases for each relevant model.
Considering the spread of values for just the extended regions gives a much larger birthrate estimate of $\dot{N} = \left(2.43 \pm 0.03\right) \times 10^{-7}~\mathrm{yr}^{-1}$, but we do not consider this estimate reliable due to the discrepancy between the models and data in this region.
Considering both the shorter and extended RGB boundary regions gives a wider spread of values and thus a larger value of $\sigma_{\dot{N}}$ in addition to a larger value of $\dot{N}_0$ compared to considering just the shorter regions, yielding an estimate of $\dot{N} = \left(2.31 \pm 0.07\right) \times 10^{-7}~\mathrm{yr}^{-1}$.

This tendency to shift the estimate of the birthrate to larger values when the RGB boundary region is extended to include the red-giant bump is worth keeping in mind when considering the results of this work.
However, including the estimates from the extended RGB regions likely overestimates the uncertainty in the birthrate.
It is important to have a tight prior on the birthrate for the unbinned likelihood analysis, otherwise the birthrate will tend to simply be adjusted so that the number of white dwarfs predicted by the likelihood function matches the total number of white dwarfs observed in the data space, which effectively prioritises fitting the normalisation constant rather than the morphology of the cooling curve and is weighted towards the faintest end of the cooling sequence where there are the most white dwarfs but also the poorest completeness.
We consider the birthrate estimate given by just the shorter RGB regions that exclude the red-giant bump to be the most accurate and thus use the result $\dot{N} = \left(2.21 \pm 0.04\right) \times 10^{-7}~\mathrm{yr}^{-1}$ given by those calculations as the prior for the birthrate in the unbinned likelihood analysis of \cref{sec:deepacs_unbinned_likelihood}.

\section{Unbinned Likelihood Analysis} \label{sec:deepacs_unbinned_likelihood}

We perform an unbinned likelihood analysis similar to that of \citet{Goldsbury2016}.
As the data used in our analysis consists of observations far from the centre of 47 Tuc, where the density profile depends very little on the distance $R$ from the cluster centre, we take the density profile to be uniform.
The distribution of photometric errors is also approximately independent of $R$ for our data, so the number density distribution function (and thus likelihood) does not depend on $R$ for our analysis.

For the analysis performed in this work, the number density distribution function $f$ for the magnitudes $m_1 = \text{F606W}$ and $m_2 = \text{F814W}$ is given by the expression
\begin{align}
\begin{split}
    f(m_1, m_2; \, \theta) = 
    &\, \dot{N} \, \fcr\left(m_1\right) \, 
    \int_{-\infty}^{\infty} \int_{-\infty}^{\infty}
    f_M\left(m_1^\prime, m_2^\prime; \, \theta_M\right)\\
    &\times 
    E\left(m_1-m_1^\prime, m_2-m_2^\prime; \, m_1^\prime, m_2^\prime\right)
    \, \mathrm{d}m_1^\prime \, \mathrm{d}m_2^\prime,
\end{split}
\label{eq:47tuc_deepacs_dist_func}
\end{align}
where $\theta$ denotes the set of parameters that the full model depends on, $\theta_M$ is the set of all model parameters except the birthrate $\dot{N}$, $E$ is the photometric error distribution function, and the quantity $\dot{N} \, f_M$ is the theoretical number density distribution function given by the model before accounting for photometric errors. 
The function $f_M$ gives the rate of change of the cooling time with respect to magnitude and depends only on the subset of parameters $\theta_M$.
The completeness correction factor $\fcr$ has been inserted into the expression for $f(m_1, m_2; \theta)$ given by \citet{Goldsbury2016} to account for the additional cleaning procedure performed in our work.
Note that $\fcr$ is a function of the magnitude $m_1$ after accounting for photometric errors.
The primed magnitudes $m_1^\prime$ and $m_2^\prime$ are the magnitude values before accounting for photometric errors, while the magnitudes $m_1$ and $m_2$ without a prime symbol are the magnitude values after accounting for photometric errors.

In the notation of \citet{Goldsbury2016}, the expression giving $f_M$ would be written as $f_M = \frac{\mathrm{d}t}{\mathrm{d}m_1^\prime \mathrm{d}m_2^\prime}$ for a cooling time $\mathrm{d}t$ over a cell of volume $\mathrm{d}m_1^\prime \mathrm{d} m_2^\prime$ in magnitude-magnitude space.
Since $m_1^\prime$ and $m_2^\prime$ are dependent variables, the expression for $f_M$ is written more formally as
\begin{equation}
    f_M\left(m_1^\prime, m_2^\prime; \, \theta_M\right)
    = \frac{\mathrm{d}t}{\mathrm{d}m_1^\prime} \, \delta\left[m_2^\prime - m_{2,\mathrm{mod}}\left(m_1^\prime; \, \theta_M\right)\right],
    \label{eq:fM}
\end{equation}
where $m_{2,\mathrm{mod}}\left(m_1^\prime; \, \theta_M \right)$ is the function that relates $m_2^\prime$ to $m_1^\prime$ for a particular cooling model (parametrized by $\theta_M$).
The set of parameters represented by $\theta_M$ consists of the parameters of the theoretical cooling models (i.e. $\mwd$ and $\lqh$), as well as the distance $d$ to 47 Tuc (from Earth) and the colour excess $E(B-V)$ due to reddening.
The full set of parameters represented by $\theta$ additionally includes the white dwarf birthrate, such that $\theta = \left\lbrace \dot{N}, \, \theta_M \right\rbrace$.

The parameters $d$ and $E(B-V)$ are needed to move the models from theory space to data space, as they are used to determine the magnitudes $m_1^\prime$ and $m_2^\prime$ predicted by a model (before accounting for photometric errors).
Since the publication of \citet{Goldsbury2016}, \gaia\ observations have been used to determine the distance to 47 Tuc to much higher accuracy, $d = 4.45 \pm 0.13~\mathrm{kpc}$ \citep{Chen2018},
than was known at the time the work of \citet{Goldsbury2016} was done, so there is less concern about its value in our work. Likewise, varying $E(B-V)$ over the very limited range of allowable values for 47 Tuc, $E(B-V) = 0.04 \pm 0.02$ \citep{1996AJ....112.1487H}, would have little effect on our results while increasing the computational cost of the analysis (if added as an additional axis of the parameter grid).
We thus keep $d$ and $E(B-V)$ fixed at values of $d = 4.45~\mathrm{kpc}$ \citep{Chen2018} and $E(B-V) = 0.04$ \citep{1996AJ....112.1487H} in our analysis.
The predicted magnitudes are calculated from the relevant theory-space model variables using bolometric corrections and these $d$ and $E(B-V)$ values.

The apparent magnitude predicted by a model in the $i$th filter is thus
\begin{equation}
    m_i^\prime = M_\mathrm{bol} - \mathrm{BC}_i + \mu + A_i,
    \label{eq:47tuc_deepacs_apparent_magnitude}
\end{equation}
where $M_\mathrm{bol}$ is the bolometric magnitude, $\mathrm{BC}_i$ is the bolometric correction for the filter labelled by index $i$, $\mu$ is the distance modulus, and $A_i$ is the extinction in the $i$th filter due to interstellar reddening, which depends on $E(B-V)$ through the extinction law. For a fixed metallicity, $\mathrm{BC}_i$ is a function of the effective temperature $T_\mathrm{eff}$ and surface gravity $g$, which the white dwarf cooling models give as a function of cooling age.
We take the distance to be $d = 4.45~\mathrm{kpc}$ \citep{Chen2018}, which gives a distance modulus of $\mu = 13.24$, and we take the colour excess to be $E(B-V) = 0.04$ \citep{1996AJ....112.1487H}.
The extinctions in the filters F606W and F814W were determined using the extinction law\footnote{The extinction values were retrieved using \url{http://stev.oapd.inaf.it/cgi-bin/cmd_3.7}, which uses this extinction law.} of \citet{1989ApJ...345..245C} and \citet{1994ApJ...422..158O} with a total $V$-band extinction of $A_V = 0.124$ and relative visibility of $R_V = 3.1$, which is typical of the Milky Way. 
The relative visibility relates the total extinction to the colour excess through the expression $A_V = R_V \, E(B-V)$.
The resultant values for the extinctions $A_i$ are
\begin{align}
    A_\mathrm{F606W} &= 0.90328 \, A_V,\\
    A_\mathrm{F814W} &= 0.59696 \, A_V.
\end{align}
We use bolometric corrections calculated by the procedure described in \citet{2006AJ....132.1221H}, which is an extension of the earlier work of \citet{1995PASP..107.1047B}, for pure-hydrogen models and the relevant HST filters\footnote{Tables with bolometric and absolute magnitudes for a number of photometric systems calculated using the same atmosphere models are available at \url{https://www.astro.umontreal.ca/~bergeron/CoolingModels/}. Absolute magnitudes in the HST filters relevant for this work were provided by Pierre Bergeron upon request.}.
These bolometric corrections for DA white dwarfs were calculated using the models of \citet{2018ApJ...863..184B} at low temperatures $T_\mathrm{eff} < 5,000~\mathrm{K}$, the models of \citet{2020ApJ...901...93B} at high temperatures $T_\mathrm{eff} > 30,000~\mathrm{K}$, and the models of \citet{2011ApJ...730..128T} at intermediate temperatures. They also incorporate the Lyman alpha profile calculations of \citet{2006ApJ...651L.137K}.

With the procedure established for calculating the distribution function given by \cref{eq:47tuc_deepacs_dist_func}, the unbinned likelihood can then be calculated from the distribution function.
The unbinned likelihood $\mathcal{L}$ is given by the expression%
\footnote{As was done in \citet{Goldsbury2016}, an additive constant that does not depend on the model parameters has been dropped from the expression for $\ln \mathcal{L}$. See \citet{Goldsbury2016} for a full derivation of this expression.}
\begin{equation}
    \ln \mathcal{L}\left(\theta\right) = \sum_i \ln f\left(m_{1i}, m_{2i}; \, \theta\right) - N_\mathrm{pred}\left(\theta\right),
\end{equation}
where $\ln \mathcal{L}$ is the natural logarithm of the likelihood, $i$ is an index that enumerates the data points (i.e. the white dwarfs observed within the boundaries defining the data space), $f(m_{1i}, m_{2i}; \, \theta)$ is the number density distribution function evaluated at the magnitude values $m_{1i}$ and $m_{2i}$ of the $i$th data point, and $N_\mathrm{pred}$ is the total number of white dwarfs predicted by the model to be in the data space. 
$N_\mathrm{pred}$ is calculated by integrating $f(m_1,m_2; \, \theta)$ over the whole data space,
\begin{equation}
    N_\mathrm{pred}\left(\theta\right) = \iint\limits_{\text{data space}} f\left(m_1, m_2; \, \theta\right) \, \mathrm{d}m_1 \, \mathrm{d}m_2.
\end{equation}
The data space used for this analysis is a truncated version of the white dwarf data space shown in \cref{fig:sharpcal_pmcal_cmd} that ends at a horizontal lower limit of $\mathrm{F606W} = 28.5$ instead of extending to the limit of $\mathrm{F606W} = 29.0$ used in the cleaning procedure.
This truncated data space region is used to ensure that the completeness (as a function of F606W magnitude) remains reasonable even after applying the completeness correction factor.
To extend the analysis to larger magnitudes, there is a trade-off between the improved statistical power of having more objects vs the reduced completeness and increased spread of the error distribution at larger magnitudes.
Multiple cut-offs for the data space between F606W of 28.0 and 29.0 were tested, and it was found that a cut-off of 28.5 optimised this trade-off while still extending to a large enough magnitude to capture the relevant feature in the cumulative luminosity function associated with convective coupling.

\begin{figure}
    \centering
    \includegraphics[trim={2.5mm 0 0 2.5mm},clip]{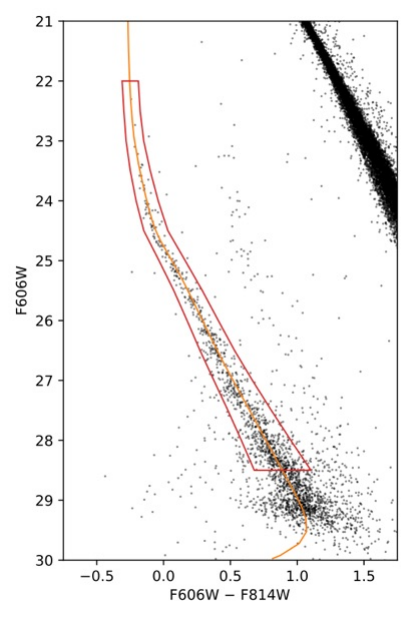}
    \caption{Data space used in unbinned likelihood analysis.
    The boundaries of the data space are indicated by the solid red curves.
    A reference cooling model in terms of input magnitudes (before accounting for photometric errors) is shown as the solid orange curve passing through the data space.
    The data space boundaries and model are shown overlaid on the cleaned HST data (points).}
    \label{fig:deepacs_dataspace}
\end{figure}

The data space ultimately used in the unbinned likelihood analysis is shown in \cref{fig:deepacs_dataspace}. 
The data space boundaries are indicated by the solid red curves enclosing most of the 47 Tuc white dwarf cooling sequence.
For reference, a cooling model as a function of input magnitudes before accounting for the distribution of photometric errors is shown as a solid orange curve that passes through 
the middle of the data space.
The data shown in \cref{fig:deepacs_dataspace} are the HST data after applying the full data cleaning procedure described in \cref{sec:deepacs_data_cleaning}.
It can be seen in \cref{fig:deepacs_dataspace} that the reference model aligns well with the observed white dwarf cooling sequence of 47 Tuc, falling approximately along the centre of this sequence.
For the model parameter ranges considered in this work, there is little variation in the corresponding model curve in the CMD, so \cref{fig:deepacs_dataspace} looks similar for every model in our grid of cooling models.
Note the data space boundaries are shown in colour-magnitude space in \cref{fig:deepacs_dataspace} for the ease of visualisation; however, the analysis was actually performed in magnitude-magnitude space.

To maximise the log-likelihood, we first evaluate the likelihood over a grid of values for the parameters $\mwd$ and $q_H$ with the birthrate fixed to the value $\dot{N}_0$ calculated from stars leaving the main sequence. 
We then analytically solve for the birthrate that optimises the likelihood at each point on the three-parameter grid and find the combination of the other parameters that give the maximum value of the likelihood after re-scaling the birthrate. 
The birthrate is treated in this way instead of as an additional axis of the parameter grid in order to reduce the memory usage of the procedure.

Let $\dot{N}_0$ be the fixed birthrate used to calculate the likelihoods on the three-parameter grid described above, and let $\ln \mathcal{L}_0$ and $N_{\mathrm{pred},0}$ be the corresponding (natural) log-likelihood and total predicted number of white dwarfs in the data space. The likelihood for this fixed birthrate is given by
\begin{equation}
    \ln \mathcal{L}_0 = \sum_{i=1}^{N_{\mathrm{obs}}} \ln f_{0,i} - N_{\mathrm{pred},0},
    \label{eq:lnL0}
\end{equation}
where $N_\mathrm{obs}$ is the number of white dwarfs actually observed within the data space and $f_{0,i}$ is the distribution function evaluated at the magnitude values of the $i$th data point with a birthrate of $\dot{N}_0$.
Let $a$ be a factor that re-scales the birthrate such that
\begin{equation}
    \dot{N} = a \ \dot{N}_0 .
    \label{eq:Ndot_rescaled}
\end{equation}
We assign a Gaussian prior with a standard deviation of $\sigma_{\dot{N}}$ to the birthrate such that $\dot{N} \sim \mathcal{N}\left(\dot{N}_0, \ \sigma_{\dot{N}} \right)$ and account for this prior in the expression for $\ln \mathcal{L}$, resulting in the expression
\begin{align}
\begin{split}
    \ln \mathcal{L} 
    = &\, \ln \mathcal{L}_0 + N_{\mathrm{obs}} \ \ln\left( a \right) - N_{\mathrm{pred},0} \left( a - 1 \right)\\
    &- \frac{1}{2} \left( \frac{\dot{N}_0}{\sigma_{\dot{N}}} \right)^2 \left(a - 1 \right)^2.
\end{split}
\label{eq:full_lnL}
\end{align}
The values of $\dot{N}_0$ and $\sigma_{\dot{N}}$ that parametrize the prior distribution for $\dot{N}$ were determined using the procedure described in \cref{sec:deepacs_birthrate}.
Note that if the priors for the other parameters represented by $\theta$ (aside from the birthrate) are taken to be uniform, then the logarithm of the joint posterior probability distribution for the parameters is equal to $\ln \mathcal{L}$ as given by \cref{eq:full_lnL} up to an additive constant (which includes the uniform prior distributions and an overall normalisation term).

The expression for $\ln \mathcal{L}$ given by \cref{eq:full_lnL} can be optimised analytically with respect to the re-scale factor $a$ without needing to evaluate the distribution function \cref{eq:47tuc_deepacs_dist_func} for different birthrate values, which reduces the computational cost of the analysis.
The extremum values for $\ln \mathcal{L}$ with respect to the birthrate re-scale factor occur when
\begin{equation}
    a = \frac{1}{2}\left( 1 - b \right) \pm \frac{1}{2} \sqrt{ \left( 1 - b \right)^2 + 4 \, b \, \frac{N_{\mathrm{obs}}}{N_{\mathrm{pred},0}} } ,
    \label{eq:optimal_rescale_factor}
\end{equation}
where the constant $b$ has been defined to be
\begin{equation}
    b \equiv N_{\mathrm{pred},0} \left( \frac{\sigma_{\dot{N}}}{\dot{N}_0} \right)^2 .
\end{equation}
Since the white dwarf birthrate (and thus the re-scale factor) must be a positive value, only the case for which $\pm$ is positive in \cref{eq:optimal_rescale_factor} is physical.

To find the combination of parameters that maximises the likelihood, i.e. the maximum likelihood estimates of the parameters, we first calculate $\ln \mathcal{L}_0$ at each point on the three-parameter grid as given by \cref{eq:lnL0}. 
At each point on this parameter grid, we then calculate $a$ as given by \cref{eq:optimal_rescale_factor} and use the result to calculate $\ln \mathcal{L}$ as given by \cref{eq:full_lnL}. 
Finally, we find the combination of parameters on the parameter grid that maximises this new $\ln \mathcal{L}$, and we get the corresponding maximum likelihood estimate of the birthrate from \cref{eq:Ndot_rescaled} using the optimal value of the re-scale factor for that parameter grid point.

The results of the unbinned likelihood analysis are presented in \cref{sec:deepacs_results}, with the results of the procedure to find the optimal parameters using the likelihood re-scaling technique described above presented in \cref{sec:deepacs_likedists}. The optimal model found by this procedure is compared to the data in \cref{sec:deepacs_cumdists} by comparing the predicted and empirical (inverse) cumulative luminosity functions.

\section{Results} \label{sec:deepacs_results}

\subsection{Likelihood Distribution} \label{sec:deepacs_likedists}

\begin{figure}
    \centering
    \includegraphics[width=0.45\textwidth]{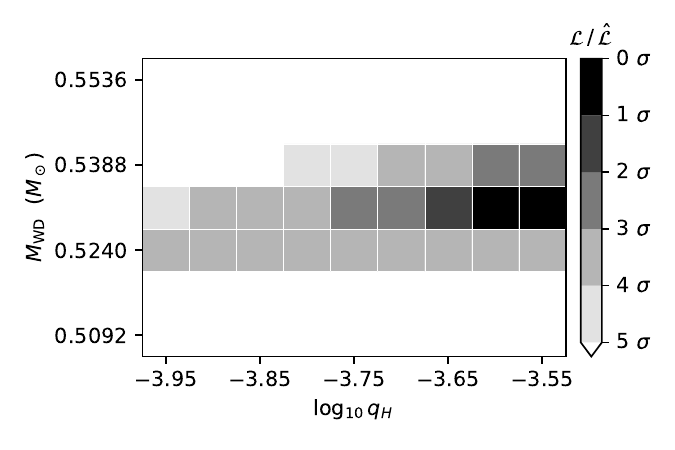}
    \caption{Likelihood (including birthrate prior) locally maximised with respect to birthrate at each location on the parameter grid.
    The likelihood has been scaled by its global maximum across the parameter grid, with contours drawn at the values of the analogously scaled probability density of a two-dimensional normal distribution evaluated at 1 to 5~$\sigma$ (darkest to lightest).}
    \label{fig:47tuc_deepacs_likelihood}
\end{figure}

The distribution of likelihood values after locally optimising the birthrate at each point on the cooling model parameter grid is shown in \cref{fig:47tuc_deepacs_likelihood}.
The quantity $\mathcal{L}$ plotted in \cref{fig:47tuc_deepacs_likelihood} is really the likelihood including the birthrate prior after re-scaling the birthrate, i.e. $\mathcal{L}$ as given by \cref{eq:full_lnL}, evaluated at each point on the parameter grid using the value of the birthrate re-scale factor that maximises $\mathcal{L}$ for that combination of cooling model parameter values.
To make the significance of the likelihood values clear, $\mathcal{L}$ has been scaled by the value $\Lhat$ of its global maximum across the entire parameter grid, and the distribution of $\mathcal{L}/\Lhat$ values is shown as a filled contour plot with the levels corresponding to the $\sigma$-level values of the similarly scaled probability density for a two-dimensional (spherically symmetric) normal distribution.
Letting $p$ denote the probability density of a two-dimensional normal distribution and $\phat$ denote the maximum value of $p$, the value of the scaled probability density corresponding to the level $n_\sigma \, \sigma$ is simply given by $\phat/p = \exp\left(-0.5 \, n_\sigma^2\right)$.
More specifically, the contour levels corresponding to 1, 2, 3, 4, and 5~$\sigma$ in \cref{fig:47tuc_deepacs_likelihood} are drawn at values of $-0.5$, $-2.0$, $-4.5$, $-8.0$, and $-12.5$ (from darkest to lightest region).
It should also be noted that \cref{fig:47tuc_deepacs_likelihood} shows only the most likely $\lqh$ values, corresponding to very thick envelopes with $\lqh \geq -4$.
However, thinner envelopes were also considered in the full analysis, with simulations performed for smaller $\lqh$ values down to values below $\lqh = -5$.

\begin{figure}
    \centering
    \includegraphics[width=0.45\textwidth]{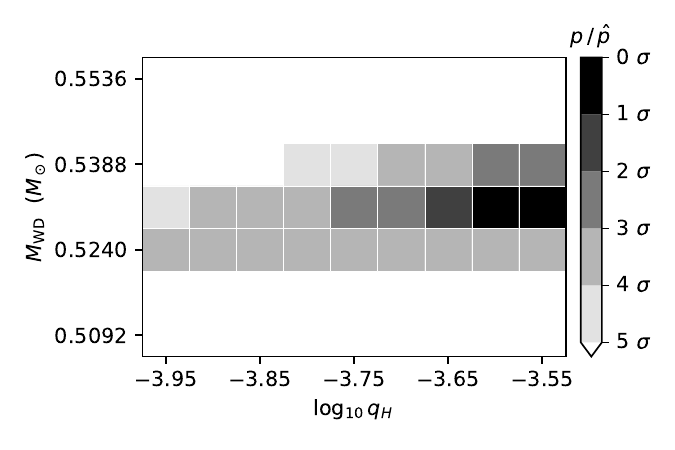}
    \caption{Joint posterior probability density distribution after marginalising over the birthrates.
    The probability density $p$ has been scaled by its maximum value $\hat{p}$ so that $p / \hat{p}$ is shown.
    The filled contours are drawn at level values corresponding to the $\sigma$ levels for a two-dimensional normal distribution.
    From darkest to lightest, these levels correspond to 1, 2, 3, 4, and 5~$\sigma$.}
    \label{fig:47tuc_deepacs_marginal_posterior}
\end{figure}

It should be emphasised that the filled contours in \cref{fig:47tuc_deepacs_likelihood} indicate ranges of probability density values, not regions of enclosed probabilities, so they are not credible regions.
The distribution shown in \cref{fig:47tuc_deepacs_likelihood} is also not quite the joint posterior distribution; rather, it is a distribution of the local maximum of the posterior with respect to birthrate at each point on the parameter grid.
Since the prior for the birthrate is a narrow Gaussian, this is expected to be very similar to the posterior distribution after marginalising over the birthrate.
This marginal posterior is shown in \cref{fig:47tuc_deepacs_marginal_posterior}, which confirms the expected similarity of these two distributions.
\cref{fig:47tuc_deepacs_marginal_posterior} is analogous to \cref{fig:47tuc_deepacs_likelihood}, with contours drawn at the same levels, but for the marginal posterior density distribution $p$ instead of $\mathcal{L}$.
Like \cref{fig:47tuc_deepacs_likelihood}, the filled contours in \cref{fig:47tuc_deepacs_marginal_posterior} are also not credible regions, though the proper credible regions would be expected to be similar for a posterior distribution that is approximately a normal distribution in the remaining parameters.

Though the quantity plotted in \cref{fig:47tuc_deepacs_likelihood} is not quite the posterior distribution, the location of its maximum value on the parameter grid is the same as the location at which the posterior probability density distribution is maximised (and the global maximum $\Lhat$ is equal to the maximum value of the full joint posterior distribution).
We find that $\Lhat$ corresponds to parameter values of $M_\mathrm{WD} = 0.5314~M_\odot$ and $\log_{10} q_H = -3.55$ (corresponding to $q_H = 2.82 \times 10^{-4}$) on the parameter grid and a birthrate of $\dot{N} = 2.27 \times 10^{-7}~\mathrm{yrs}^{-1}$ after re-scaling.
The corresponding value of the log-likelihood is $\ln \Lhat = 5385.35$.
These maximum likelihood estimate results are summarised in \cref{tab:deepacs_MLE_results}.

\begin{table}
    \centering
    \caption{Maximum likelihood estimates of parameters and corresponding log-likelihood value from unbinned likelihood analysis.
    These are the parameter values and likelihood of the best-fitting model.}
    \begin{tabularx}{\columnwidth}{>{\centering\arraybackslash}X >{\centering\arraybackslash}X}
        \toprule
        Parameter & Value \\
        \midrule
        $M_\mathrm{WD}$ & $0.5314~M_\odot$ \\
        $\log_{10} q_H$ & $-3.55$ \\
        $\dot{N}$ & $2.27 \times 10^{-7}~\mathrm{yrs}^{-1}$ \\
        \midrule
        $\ln \Lhat$ & $5385.35$ \\
        \bottomrule
    \end{tabularx}
    \label{tab:deepacs_MLE_results}
\end{table}

\subsection{Best-Fitting Model} \label{sec:deepacs_cumdists}

\begin{figure}
    \centering
    \includegraphics[width=\columnwidth]{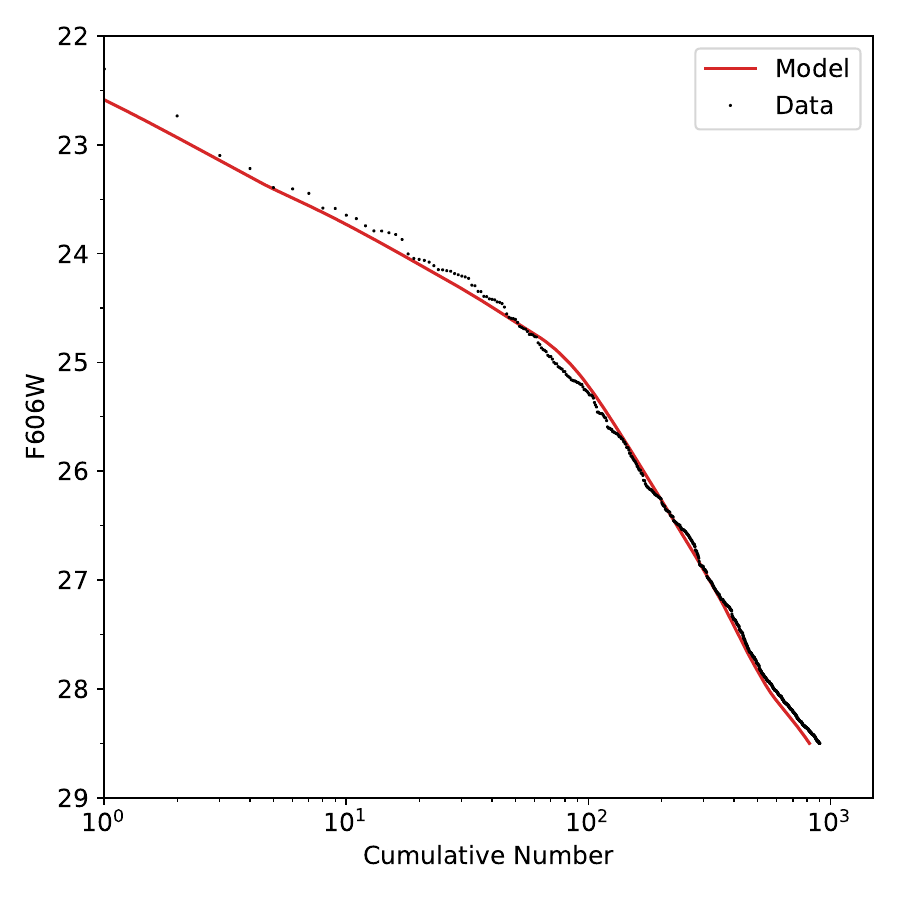}
    \caption{Inverse cumulative luminosity function for F606W magnitude of data (black points) compared to optimal model determined by the unbinned likelihood analysis (red curve).}
    \label{fig:deepacs_best_model_cumnum_f606w}
\end{figure}

\begin{figure}
    \centering
    \includegraphics[width=\columnwidth]{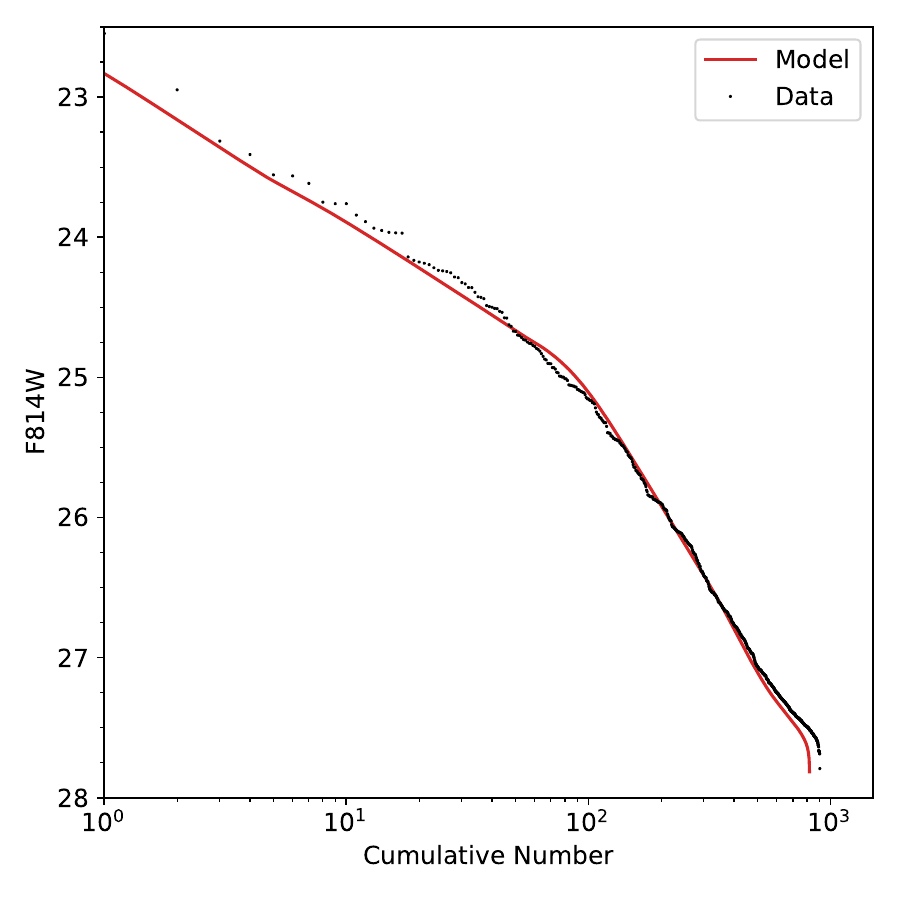}
    \caption{Inverse cumulative luminosity function for F814W magnitude of data (black points) compared to optimal model determined by the unbinned likelihood analysis (red curve).}
    \label{fig:deepacs_best_model_cumnum_f814w}
\end{figure}

The inverse cumulative luminosity function, i.e. the inverse of the cumulative number distribution of white dwarfs in the data space as a function of magnitude, is shown in \cref{fig:deepacs_best_model_cumnum_f606w,fig:deepacs_best_model_cumnum_f814w} for the best-fitting model from the unbinned likelihood analysis (red curve) in comparison to the empirical distribution for the HST data (black points) that the models were fitted to.
The cumulative number as a function of magnitude for the model is given by integrating the number density distribution function,
\begin{equation}
    N(m_1; \, \theta) = \int_{-\infty}^{m_1} \mathrm{d} m_1^* \int_{-\infty}^{\infty} \mathrm{d} m_2 \ f(m_1^*, m_2; \theta)
    \label{eq:cumnum}
\end{equation}
where $N$ is the (predicted) number of white dwarfs in the data space with magnitude $\leq m_1$ in filter 1 (in this case F606W) and $f(m_1, m_2; \theta)$ is the number density distribution function given by \cref{eq:47tuc_deepacs_dist_func} evaluated over the $(m_1, m_2)$ coordinates of the data space and equal to zero outside of the data space.
An equivalent expression for the cumulative number distribution as a function of $m_2$ can be written by swapping the indices 1 and 2 in \cref{eq:cumnum}.

Note that the cumulative number distribution is proportional to the white dwarf birthrate through the proportional dependence of $f(m_1,m_2)$ on the birthrate; a larger birthrate would shift the whole model distribution to the right in \cref{fig:deepacs_best_model_cumnum_f606w,fig:deepacs_best_model_cumnum_f814w}, while a smaller birthrate would shift it to the left.
The other model parameters that were varied in the analysis affect the morphology of the distribution.
The cumulative number distribution for the data is given by simply counting a list of the data points ordered by magnitude.
The incompleteness of the data is accounted for in the distribution functions of the models, so it does not need to be corrected for in plotting the distribution of the data in \cref{fig:deepacs_best_model_cumnum_f606w,fig:deepacs_best_model_cumnum_f814w}.

It can be seen from \cref{fig:deepacs_best_model_cumnum_f606w,fig:deepacs_best_model_cumnum_f814w} that the optimal model from the unbinned likelihood analysis well reproduces the empirical distribution for most of the white dwarf data space, which is indicative of a good fit to the data.
As a simple quantitative check of the goodness-of-fit of the optimal model to the data, we performed a one-sample KS test for the cumulative marginal probability distribution with respect to each magnitude (after integrating over the other magnitude).
Each KS tests compares the cumulative probability density function predicted by the model for a given magnitude to the empirical cumulative fraction as a function of that magnitude.
The \textit{p}-value returned by the one-sample KS test gives the probability of drawing a sample from the model distribution for which the corresponding sample distribution differs from the model distribution by at least as much as the observed empirical distribution differs from the model distribution.
Very small \textit{p}-values indicate a low probability that the data was drawn from the model distribution and thus a poor fit, while larger \textit{p}-values are expected if the data was drawn from the model distribution (i.e. if the model distribution and empirical distribution are equivalent) and thus indicate a good fit.
The \textit{p}-values from these KS tests are given in \cref{tab:47tuc_deepacs_KS_test_results}.
The large \textit{p}-values (well above a reasonable threshold of $10^{-4}$) in both cases indicate a good fit.

\begin{table}
    \centering
    \caption{Results of KS tests comparing the one-dimensional marginal cumulative probability distribution functions predicted by the optimal model to the corresponding empirical distribution.}
    \begin{tabularx}{\columnwidth}{>{\centering\arraybackslash}X >{\centering\arraybackslash}X}
    \toprule
        Magnitude Variable & \textit{p}-value \\
    \midrule
        F606W & 0.077 \\
        F814W & 0.041 \\
    \bottomrule
    \end{tabularx}
    \label{tab:47tuc_deepacs_KS_test_results}
\end{table}

Note that the distributions compared by each KS test are cumulative fractions normalized to unity over the magnitude range, rather than the cumulative number distributions as shown in \cref{fig:deepacs_best_model_cumnum_f606w,fig:deepacs_best_model_cumnum_f814w}.
These KS tests specifically assess the morphology of the (inverse) cumulative luminosity function with respect to a particular magnitude for the optimal combination of cooling model parameters ($M_\mathrm{WD}$, $q_H$).
The birthrate parameter simply re-scales the total number of white dwarfs predicted by the model (i.e. the normalization of the number density distribution), but the KS test is applied to the probability distribution function (normalized to unity).
Thus, while the birthrate was important in determining the optimal cooling model using the unbinned likelihood analysis, it does not directly affect the morphology assessed by the KS test.
Nonetheless, we note that the birthrate found for the optimal model only deviates from the prior value by 
\begin{equation}
    \dot{N} - \dot{N}_0 = 1.5 \, \sigma_{\dot{N}},
\end{equation}
suggesting the birthrate value is reasonable.

\section{Discussion}  \label{sec:deepacs_discussion}

Previous work studying the cooling of white dwarfs in 47 Tuc such as \citet{Goldsbury2016} and \citet{Obertas2018} used cooling models based on \verb|test_suite| examples from older versions of MESA that had diffusion turned off for white dwarf evolution.
Our cooling simulations, however, followed more recent MESA examples that had diffusion turned on during white dwarf evolution.
\citet{Obertas2018} used the same data as our work and considered old white dwarfs down to faint enough magnitudes in the cooling sequence to reveal the effect of core crystallisation in the (inverse) cumulative luminosity function, but the models shown in that work were not fitted to the data. 
\citet{Goldsbury2016} found models that fitted the data well, but compared the models to data of younger white dwarfs than considered in this work.
As the focus of \citet{Goldsbury2016} was neutrino cooling, which is important at early white dwarf cooling times (and negligible at late cooling times), the data space selections used in that work stopped at much brighter magnitudes\footnote{\citet{Goldsbury2016} used HST data in different filters than our data, so the magnitude values are not directly comparable, but they can be compared through the corresponding theoretical luminosity values.} than the data space used in our analysis.
Since we are interested in the thickness of the H envelope as a key parameter in our work, rather than it being a nuisance parameter as it was in the work of \citet{Goldsbury2016}, and the effect of this parameter is best analysed using very old white dwarfs, our data space needs to extend to fainter magnitudes.

The white dwarf cooling sequence of 47 Tuc has also recently been studied by \citet{2025MNRAS.541.1390S}. 
The focus of that work was using the white dwarf luminosity function to determine the age of 47 Tuc, which was found to be $11.8 \pm 0.5~\mathrm{Gyr}$ (in agreement with age determinations based on the main-sequence turn-off and the masses and radii of two eclipsing binaries).
\citet{2025MNRAS.541.1390S} used JWST infrared photometric observations in conjunction with the same HST data used in our work, with the HST data used to determine proper motions for the purpose of removing SMC contamination from the data and used as a check of the results for the JWST data.
The JWST data were used to construct completeness-corrected empirical white dwarf luminosity functions, which were compared to theoretical white dwarf luminosity functions to determine the cluster age.
\citet{2025MNRAS.541.1390S} used white dwarf models with a fixed H envelope thickness, for which the mass of the pure H layer was $10^{-4}~M_\mathrm{WD}$, and used isochrones to construct synthetic populations of white dwarfs from which the theoretical white dwarf luminosity functions were constructed for different cluster ages.

The key feature of the white dwarf luminosity function that \citet{2025MNRAS.541.1390S} used to determine the cluster age was the bump at very faint magnitudes (corresponding to $\mathrm{F606W} \sim 29.1$ in the HST data), which, as noted by the authors, corresponds to the sudden increase in white dwarf (and progenitor) mass at those faintest magnitudes \citep[see Fig. 8 of][]{2025MNRAS.541.1390S}.
In contrast, the data space used in our analysis ends at a brighter magnitude just before this feature appears, with our analysis restricted to the regime where the white dwarfs are not very sensitive to the details of the progenitors and the mass is still approximately constant, as indicated by Fig. 8 of \citet{2025MNRAS.541.1390S} and comparing the JWST and HST white dwarf luminosity functions shown in Fig. 4 of the same work.
These different choices of data space are suited to the complementary but different purposes of the work of \citet{2025MNRAS.541.1390S} compared to our work.

The focus of our work was on constraining the thickness of the H envelope. 
It should be noted that a few other model parameters could potentially be relevant to our results but were beyond the scope of this work to study.
\citet{2025ApJ...983..158P} recently studied the uncertainties in white dwarf cooling ages associated with the evolutionary models used to determine these ages, and that work identified both the conductive opacities and the composition profile as key sources of uncertainty in the models.

For our cooling models, we used the conductive opacities included by default in MESA version 15140 (mesa-r15140), which are the conductive opacities of \citet{2007ApJ...661.1094C}.
More recent revised calculations of conductive opacities for H and He in the regime of partial degeneracy and moderate Coulomb coupling have been performed by \citet{2020ApJ...899...46B}, though there is an ongoing debate about the accuracy of these new calculations \citep{2021A&A...654A.149C}.
The revised opacities of \citet{2020ApJ...899...46B} have been added to more recent versions of MESA, but they had not yet been implemented in the MESA version used in our work.
\citet{2025ApJ...983..158P} found that uncertainties in conductive opacities caused a larger (though comparable) range of variations in the white dwarf cooling age than did changes in the H layer thickness, although it should be noted that \citet{2025ApJ...983..158P} only consider H layer thicknesses up to an upper limit of $\log q_H = -4.00$.
In our work, we consider models with thicker envelopes than this limit and find significant variations amongst the cooling curves of these thicker-envelope models. In fact, it is models with H layers thicker than this upper limit that are found to be most important for our analysis.
Nevertheless, the uncertainties in the conductive opacities are an important source of uncertainty not accounted for in our work, and using different conductive opacities could potentially change our results.

The other key source of uncertainty noted by \citet{2025ApJ...983..158P} was the composition profile, in particular the C/O abundance profile in the core. 
For their analysis, \citet{2025ApJ...983..158P} set the composition profile of the white dwarf models directly and studied a range of different profiles.
In our work, the composition profile was set by the pre-main sequence to white dwarf evolution simulation that produced the initial white dwarf model from which the white dwarf evolution simulations were generated.
The resulting white dwarf models used in our work all had similar C/O abundance profiles and He layer thicknesses, with a typical composition profile shown in \cref{fig:composition_profile} and discussed in \cref{sec:deepacs_models_initial_model}.
Though the composition profile was well motivated by our earlier simulation, there is still an uncertainty associated with the C/O abundance profile and He layer thickness that is not accounted for in our analysis.

\section{Conclusions}  \label{sec:deepacs_conclusions}

In this work, we performed a detailed analysis of the cooling of white dwarfs to late cooling times in the globular cluster 47 Tuc using archival data of deep observations taken by HST.
These deep photometric observations resolve the white dwarf cooling sequence of 47 Tuc to faint enough magnitudes that a bump associated with the onset of convective coupling and core crystallisation can be seen in the luminosity function. This was shown by \citet{Obertas2018} using this same data, though a statistical analysis of model fits was not performed in that work.
We built upon the work of \citet{Obertas2018} by performing a detailed statistical analysis that accounted for different H envelope thickness, white dwarf mass, and white dwarf birthrate using the unbinned likelihood.

A cleaning procedure consisting of cuts in proper motion and the photometry quality-of-fit parameter \sharpparam\ were performed to remove contamination from the SMC in the 47 Tuc white dwarf cooling sequence, and this cleaning procedure was carefully calibrated to account for any potential residual contamination and the reduction of completeness due to cleaning.
This cleaning procedure and calibration was particularly important for using data at the very faint end of the cooling sequence where the phenomena of convective coupling of the envelope to the core and the crystallisation of the core begin to occur.
The thickness of the bump in the luminosity function due to convective coupling in particular is sensitive to the H envelope thickness. 
The inclusion of these very old, faint white dwarfs in the analysis is thus important to distinguish cooling models of different H envelope thickness.
Though the data for these older, fainter white dwarfs has lower completeness than younger, brighter white dwarfs, there is a much larger number of the fainter white dwarfs in the data, so including them in the analysis furthermore provides improved statistical power.

The stellar evolution software MESA was used to produce a suite of white dwarf cooling models for different values of the H envelope thickness parameter and the total mass of white dwarf.
In addition to the cooling model parameters, the full model for the number density distribution function used in the unbinned likelihood analysis was also sensitive to the white dwarf birthrate, for which a prior value was determined using \gaia\ EDR3 data of stars on the RGB.
The optimal model found by the unbinned likelihood analysis corresponded to a H envelope thickness parameter of $\lqh = -3.55$ (corresponding to $q_H = 2.82 \times 10^{-4}$), a white dwarf mass of $M_\mathrm{WD} = 0.5314~M_\odot$, and a white dwarf birthrate of $\dot{N} = 2.27 \times 10^{-7}~\mathrm{yrs}^{-1}$.
We find that this best-fitting model well reproduces the cumulative white dwarf luminosity function to magnitudes faint enough to resolve features related to the onset of convective coupling and core crystallisation, and we find overall that models with thicker H envelopes are favoured.

The analysis of the deep HST ACS/WFC data considered in this work is limited by the increasingly poor completeness with increasing magnitude, particularly at the faint end of the cooling sequence that is most important for our analysis.
Improved data with better completeness at the faintest magnitudes of the cooling sequence, as could for example be obtained with a newer telescope like JWST, could potentially provide improved bounds on the H envelope thickness in the future following the analysis procedure used in this work.
This analysis, however, has already pushed the limits of what can be learned from the data considered in this work.

\section*{Acknowledgements}

This work has been supported by the Natural Sciences and Engineering Research Council of Canada through the Discovery Grants program and Compute Canada.
This research was also supported in part through computational resources and services provided by Advanced Research Computing at the University of British Columbia.

This work has made use of data from the European Space Agency (ESA) mission \gaia\ (\url{https://www.cosmos.esa.int/gaia}), processed by the \gaia\ Data Processing and Analysis Consortium (DPAC, \url{https://www.cosmos.esa.int/web/gaia/dpac/consortium}).
Funding for the DPAC has been provided by national institutions, in particular the institutions participating in the \gaia\ Multilateral Agreement.

The main research in this work is based on observations made with the NASA/ESA \textit{Hubble Space Telescope} obtained from the Space Telescope Science Institute, which is operated by the Association of Universities for Research in Astronomy, Inc., under NASA contract NAS 5–26555. These observations are associated with proposal GO-11677 (PI: H. Richer).

We thank Pierre Bergeron for providing bolometric correction data for the HST photometric system upon request. These data correspond to the synthetic colours from the website ``Synthetic Colors and Evolutionary Sequences of Hydrogen- and Helium-Atmosphere White Dwarfs'' at \url{http://www.astro.umontreal.ca/~bergeron/CoolingModels/} but calculated for the relevant HST filters.
We retrieved the bolometric correction data for the \gaia\ filters and the extinction data for all filters from the \texttt{PARSEC} database through the website \url{https://stev.oapd.inaf.it}.

\section*{Data Availability}

The HST data used in this work are publicly available through the Mikulski Archive for Space Telescopes (MAST, \url{https://archive.stsci.edu/}). 
The images in each filter for visits consisting of the combined exposures from 5 orbits (except for Visit 24, which consists of 6 orbits) can be accessed through the MAST Portal at \url{https://mast.stsci.edu/portal/Mashup/Clients/Mast/Portal.html}.
The final stacked FITS mosaics combining all orbits, the source catalogue, and the artificial stars data are all available as MAST high level science products at \url{https://archive.stsci.edu/prepds/deep47tuc/}.
The \gaia\ data used in this work are available through TAP Vizier and the \gaia\ archive.
The \texttt{PARSEC} bolometric correction and extinction data used in this work are available at \url{https://stev.oapd.inaf.it}.

%%%%%%%%%%%%%%%%%%%%%%%%%%%%%%%%%%%%%%%%%%%%%%%%%%

%%%%%%%%%%%%%%%%%%%% REFERENCES %%%%%%%%%%%%%%%%%%

\bibliographystyle{mnras}
\bibliography{main}

%%%%%%%%%%%%%%%%%%%%%%%%%%%%%%%%%%%%%%%%%%%%%%%%%%

%%%%%%%%%%%%%%%%% APPENDICES %%%%%%%%%%%%%%%%%%%%%

\appendix
\section{SMC Contamination} \label[appendix]{sec:appendix_smc_contamination}

Let the subscript ``D'' indicate the full \sharpparam-cleaned dataset before any proper motion cuts or CMD cuts, and let the subscripts ``S'' and ``T'' denote the subsets of the true SMC stars and the true 47 Tuc stars, respectively, within the full dataset.
If a proper motion cut has been applied to one of these population (``D'', ``S'', or ``T''), that will be denoted with the subscript ``ps'' for the SMC proper motion cut and ``pt'' for the 47 Tuc proper motion cut.
If a CMD cut has been applied, that will be denoted with the subscript ``cs'' for the SMC CMD cut, ``cw'' for the 47 Tuc white dwarf CMD cut, and ``cm'' for the 47 Tuc main-sequence CMD cut.

The 47 Tuc white dwarf sample used in the main analysis is selected by applying both the 47 Tuc proper motion cut and the 47 Tuc CMD cut, and the number of stars in this sample is $N_\mathrm{D,pt,cw}$
Assuming SMC stars are the only contaminants in this sample after \sharpparam\ cleaning, then
\begin{equation}
    N_\mathrm{D,pt,cw} = N_\mathrm{T,pt,cw} + N_\mathrm{S,pt,cw},
\end{equation}
where $N_\mathrm{T,pt,cw}$ is the true number of 47 Tuc white dwarfs in the sample and $N_\mathrm{S,pt,cw}$ is the true number of SMC contaminants in the sample.
We want to estimate $N_\mathrm{S,pt,cw}$ using numbers that can actually be calculated by applying cuts to the full dataset.

If the SMC proper motion cut is chosen such that all of the objects selected by this cut, when applied to the full dataset, are actually SMC stars, then
\begin{alignat}{2}
    &N_\mathrm{S,ps,cs} &&= N_\mathrm{D,ps,cs},\label{eq:pscs}\\
    &N_\mathrm{S,ps,cw} &&= N_\mathrm{D,ps,cw},\label{eq:pscw}
\end{alignat}
where $N_\mathrm{S,ps,cs}$ is the number of objects selected when both the SMC proper motion cut and SMC CMD cut are applied to the true SMC stars, $N_\mathrm{D,ps,cs}$ is the number of objects selected when both the SMC proper motion cut and SMC CMD cut are applied to the full dataset, $N_\mathrm{S,ps,cw}$ is the number of objects selected when both the SMC proper motion cut and 47 Tuc white dwarf CMD cut are applied to the true SMC stars, and $N_\mathrm{D,ps,cw}$ is the number of objects selected when both the SMC proper motion cut and 47 Tuc white dwarf CMD cut are applied to the full dataset.

Likewise, if the SMC CMD cut is chosen such that all of the objects selected by this cut, when applied to the full dataset, are actually SMC stars, then
\begin{equation}
    N_\mathrm{S,pt,cs} = N_\mathrm{D,pt,cs},\label{eq:ptcs}
\end{equation}
where $N_\mathrm{S,pt,cs}$ is the number of objects selected when both the 47 Tuc proper motion cut and SMC CMD cut are applied to the full dataset and $N_\mathrm{D,pt,cs}$ is the number of objects selected when both the 47 Tuc proper motion cut and SMC CMD cut are applied to the true SMC stars.

In practice, \cref{eq:pscs,eq:pscw,eq:ptcs} are only approximately true, as some 47 Tuc stars could potentially survive the relevant cuts, particularly the SMC CMD cut in combination with the 47 Tuc proper motion. 
This makes $N_\mathrm{D,pt,cs}$ in particular an upper limit on $N_\mathrm{S,pt,cs}$, which will translate to our estimate of $N_\mathrm{S,pt,cw}$ really being an upper limit on $N_\mathrm{S,pt,cw}$.
However, the SMC cuts are chosen such that \cref{eq:pscs,eq:pscw,eq:ptcs} are good approximations.

Assuming the ratio of SMC stars that survive the 47 Tuc white dwarf CMD cut to SMC stars that survive the SMC CMD cut is the same for both the 47 Tuc and SMC proper motion cuts, as it should be, then
\begin{align}
    N_\mathrm{S,pt,cw}
    &= N_\mathrm{S,pt,cs} \, \frac{N_\mathrm{S,ps,cw}}{N_\mathrm{S,ps,cs}}\label{eq:Sptcw_exact}\\
    &\approx N_\mathrm{D,pt,cs} \, \frac{N_\mathrm{D,ps,cw}}{N_\mathrm{D,ps,cs}}.\label{eq:Sptcw_approx}
\end{align}
The expression on the right-hand side of \cref{eq:Sptcw_approx} is equivalent to the definition of $\Ncontam$ in \cref{eq:Ncontam}, used in \cref{sec:deepacs_pmcal} to estimate the number of SMC contaminants in the 47 Tuc white dwarf data space.

Note that the numbers corresponding to samples selected using a cut designed to select 47 Tuc stars, i.e. $N_\mathrm{D,pt,cs}$ (47 Tuc proper motion cut) and $N_\mathrm{D,ps,cw}$ (47 Tuc white dwarf CMD cut), both occur in the numerator. 
These are the numbers for which 47 Tuc stars are most likely to be miscounted as SMC stars. 
Whereas $N_\mathrm{D,ps,cs}$, which appears in the denominator of \cref{eq:Sptcw_approx}, is calculated from the sample that is least likely to contain any misclassified 47 Tuc stars as it is selected by applying both the SMC proper motion cut and the SMC CMD cut.
Thus, \cref{eq:Sptcw_approx} is really an upper limit on $N_\mathrm{S,pt,cw}$, though this upper limit should also be close to the value of $N_\mathrm{S,pt,cw}$.
As the value of $\Ncontam$ is found in \cref{sec:deepacs_pmcal} to be small at all magnitudes of interest, knowing the upper limit on $N_\mathrm{S,pt,cw}$ is sufficient for our purpose of determining that the number of SMC contaminants in the 47 Tuc white dwarf data space after proper motion cleaning is small enough to be neglected.

%%%%%%%%%%%%%%%%%%%%%%%%%%%%%%%%%%%%%%%%%%%%%%%%%%

% Don't change these lines
\bsp	% typesetting comment
\label{lastpage}
\end{document}